\documentclass[fleqn,usenatbib]{mnras}

\usepackage{kantlipsum}
\usepackage{hyperref}  

\usepackage{etoolbox}
\makeatletter
\newcount\c@additionalboxlevel
\newcount\c@maxboxlevel
\patchcmd\@combinedblfloats{\box\@outputbox}{%
  \stepcounter{additionalboxlevel}%
  \box\@outputbox
}{}{\errmessage{\noexpand\@combinedblfloats could not be patched}}

\AtBeginShipout{%
  \ifnum\value{additionalboxlevel}>\value{maxboxlevel}%
    \typeout{Warning: maxboxlevel might be too small, increase to %
      \the\value{additionalboxlevel}%
    }%
  \fi 
  \@whilenum\value{additionalboxlevel}<\value{maxboxlevel}\do{%
    \typeout{* Additional boxing of page `\thepage'}%
    \setbox\AtBeginShipoutBox=\hbox{\copy\AtBeginShipoutBox}%
    \stepcounter{additionalboxlevel}%
  }%
  \setcounter{additionalboxlevel}{0}%
}
\makeatother

\usepackage{mathptmx}

\usepackage[T1]{fontenc}
\usepackage{ae,aecompl}

\usepackage{graphicx,epstopdf}
\usepackage{float}
\usepackage{flafter}
\usepackage{amsmath}
\usepackage{amssymb}
\usepackage{booktabs}
\usepackage{multirow}
\usepackage{hhline}
\usepackage{enumitem}
\usepackage[normalem]{ulem}
\usepackage{pdflscape}
\usepackage{longtable}
\usepackage{siunitx}
\usepackage{placeins}
\usepackage{subcaption}
\usepackage{hyperref}

\DeclareSIUnit\parsec{pc}
\DeclareSIUnit\mpc{Mpc}
\DeclareSIUnit\year{yr}
\DeclareSIUnit\Jy{Jy}

\newcommand{\gdr}{$\delta_{ \rm GDR}$}

\newcommand{\lpah}{$L_{\rm 6.2}$}
\newcommand{\ew}{${\rm EW}_{6.2}$}
\newcommand{\lpahsev}{$L_{\rm 7.7}$}
\newcommand{\lpahall}{$L_{\rm PAH}$}
 
\newcommand{\sst}{{\it Spitzer Space Telescope}}
\newcommand{\hii}{{\sc h~ii}}

\newcommand{\s}{{\it Spitzer}}
\newcommand{\h}{{\it Herschel}}
\newcommand{\msol}{$\rm M_{\odot}$}
\newcommand{\md}{$M_{\rm dust}$}
\newcommand{\sfr}{M$_{\odot}$ \si{\per\year}}

\newcommand{\lsol}{L$_{\odot}$}
\newcommand{\lir}{$L_{\rm IR}$}
\newcommand{\ms}{$M_{\ast}$}
\newcommand{\lco}{$L^{\prime}_{\rm CO}$}
\newcommand{\ulco}{\si{\kelvin\km\per\s\parsec\squared}}

\newcommand{\mol}{$M_{\rm H_{\rm 2}}$}
\newcommand{\mat}{$M_{\rm H_{\rm I}}$}

\newcommand{\aco}{$\alpha_{\rm CO}$}
\newcommand{\apah}{$\alpha_{\rm 6.2}$}
\newcommand{\uaco}{$\rm M_{\odot}/$(\si{\kelvin\km\per\s\parsec\squared})}

\newcommand{\mgas}{$M_{\rm gas}$}

\newcommand{\lmid}{$L_{\rm 8}$}
\newcommand{\Mstar}{$M_{\ast}$}

\title[PAHs as tracers of the molecular gas in galaxies]{PAHs as tracers of the molecular gas in star-forming galaxies}

\author[I. Cortzen et al.]{
I. Cortzen,$^{1,2}$
J. Garrett,$^{3}$
G. Magdis,$^{1,2}$
D. Rigopoulou,$^{3}$
F. Valentino,$^{1,2}$
\newauthor M. Pereira-Santaella,$^{3}$ F. Combes,$^{4}$ A. Alonso-Herrero,$^{5}$ S. Toft,$^{1,2}$ E. Daddi,$^{6}$
\newauthor D. Elbaz,$^{6}$ C. G\'omez-Guijarro,$^{1,2}$, M. Stockmann,$^{1,2}$ J. Huang,$^{7}$, C. Kramer$^{8}$
\\
$^{1}$Cosmic Dawn Center (DAWN), Niels Bohr Institute, University of Copenhagen, Juliane Maries Vej 30, DK-2100 Copenhagen \O \\  DTU-Space, Technical University of Denmark,  Elektrovej 327, DK-2800 Kgs.\ Lyngby\\
$^{2}$Dark Cosmology Centre, Niels Bohr Institute, University of Copenhagen, Juliane Maries Vej 30, DK-2100 Copenhagen \O\\
$^{3}$Department of Physics, University of Oxford, Keble Road, Oxford OX1 3RH, UK\\
$^{4}$Observatoire de Paris, LERMA, College de France, DNRS, PSL, Sorbonne Univ. UPMC, F-75014, Paris, France\\
$^{5}$Centro de Astrobiolog\'{\i}a (CAB, CSIC-INTA), ESAC Campus,E-28692 Villanueva de la Ca\~nada, Madrid, Spain\\
$^{6}$CEA Saclay, Laboratoire AIM-CNRS-Universit\'e Paris Diderot, Irfu/SAp, Orme des Merisiers, F-91191 Gif-sur Yvette, France\\
$^{7}$National Astronomical Observatories of China, Chinese Academy of Sciences, 100012 Beijing, PR China\\
$^{8}$Instituto de Radioastronom\'{\i}a Milim\'etrica, Av. Divina Pastora 7, N\'ucleo Central, E-18012 Granada, Spain
\\
}

\pubyear{2018}
\begin{document}
\label{firstpage}
\pagerange{\pageref{firstpage}--\pageref{lastpage}}
\maketitle
\begin{abstract}
We combine new CO(1--0) line observations of 24 intermediate redshift galaxies ($0.03 < z < 0.28$) along with literature data of galaxies at $0 < z < 4$ to explore scaling relations between the dust and gas content using PAH 6.2 \si{\micro\metre} (\lpah), CO (\lco), and infrared (\lir) luminosities for a wide range of redshifts and physical environments. Our analysis confirms the existence of a universal \lpah--\lco\ correlation followed by normal star-forming galaxies (SFGs) and starbursts (SBs) at all redshifts. This relation is also followed by local ULIRGs that appear as outliers in the \lpah--\lir\ and \lir--\lco\ relations from the sequence defined by normal SFGs.
The emerging tight ($\sigma \approx 0.26$ dex) and linear ($\alpha = 1.03$) relation between \lpah\ and \lco\ indicates a \lpah\ to molecular gas (\mol) conversion factor of $\alpha_{\rm 6.2} =$ \mol/\lpah $= (2.7\pm1.3) \times$ \aco, where \aco\ is the \lco\ to \mol\ conversion factor. We also find that on galaxy integrated scales, PAH emission is better correlated with cold rather than with warm dust emission, suggesting that PAHs are associated with the diffuse cold dust, which is another proxy for \mol. 
Focusing on normal SFGs among our sample, we employ the dust continuum emission to derive \mol\ estimates and find a constant \mol/\lpah\ ratio of $\alpha_{\rm 6.2} = 12.3$ \msol/\lsol ($\sigma\approx 0.3$\,dex). This ratio is in excellent agreement with the \lco-based \mol/\lpah\ values for \aco $= 4.5$ \uaco\ which is typical of normal SFGs. We propose that the presented \lpah --\lco\ and \lpah --\mol\ relations will serve as useful tools for the determination of the physical properties of high$-z$ SFGs, for which PAH emission will be routinely detected by the {\it James Webb Space Telescope} ({\it JWST}).

\end{abstract}




\section{Introduction}\label{sec:intro}
The mid-infrared (MIR; $3-25$ \si{\micro\metre}) spectrum of star-forming galaxies (SFGs) is dominated by strong emission features generally attributed to polycyclic aromatic hydrocarbons (PAHs) \citep{Sellgren1984, Puget1989, Helou2001, Pahre2004, Tielens2008}. The extensive observations of PAH emission in galaxies at both low and high redshifts from either the \textit{Infrared Space Observatory} \citep{Genzel1998, Lutz1998, Rigopoulou1999} or IRS on the \sst\ \citep{Armus2007, Houck2007a, Spoon2007, Valiante2007, Yan2007, Farrah2008, Sajina2008, Murphy2009, ODowd2009, Veilleux2009, Fadda2010, PereiraSantaella2010, Riechers2014} indicate that they are ubiquitous and an important tracer of the interstellar medium (ISM). PAH molecules, which are stochastically heated by optical and UV photons, dominate the photoelectric heating rates of the neutral gas and the ionization balance within molecular clouds \citep{Bakes1994}. 
The emission arising from these abundant species can contribute up to 20\% of the total infrared (IR) emission in galaxies depending on the physical conditions \citep{Smith2007, Dale2009}. Hard UV photon fields are thought to destroy, fragment or ionize the PAH molecules \citep{Boulanger1988, Boulanger1990, Helou1991, Pety2005}, whereas low-metallicity systems reveal suppressed PAH emission \citep{Engelbracht2005, Hunt2010}. The origin of PAHs has been widely discussed in previous studies suggesting that they  can be formed in either the envelopes or outflows of carbon-rich AGB stars \citep{Latter1991, Cherchneff1992}, massive red supergiants \citep{Melbourne2013} or in the ISM itself \citep{Tielens1987, Puget1989, Herbst1991, Sandstrom2010, Sandstrom2012, Sandstrom2013}. 

In the local universe, PAH emission and its link to star formation has been thoroughly studied within the Milky Way and in nearby galaxies through various star formation tracers: individual observations of \hii\ regions revealed that the PAH emission is found in shell-like structures around the star-forming regions \citep{Churchwell2006, Rho2006, SmithBrooks2007}, with a notable decrease of their strength within the \hii\ regions \citep{Helou2004, Calzetti2005, Calzetti2007, Lebouteiller2007, Povich2007, Thilker2007}. On larger scales, previous studies have found that star-forming galaxies (SFGs) at both low and high redshifts follow a linear relation between the integrated luminosity of the PAH 6.2 \si{\micro\metre} feature (\lpah) and the total infrared luminosity (\lir), where the latter is the sum of the re-radiated emission from dust grains and a commonly used tracer for the star formation rate (SFR) \citep{Schmidt1959, Kennicutt1998, Roussel2001, ForsterSchreiber2004, Armus2007, Huang2009, MenendezDelmestre2009, Rujopakarn2013}. PAH emission has also been observed in both ultra-luminous infrared galaxies (ULIRGs: \lir > $10^{12}$~\lsol) \citep{Genzel1998, Armus2007, Desai2007} and galaxies with the presence of an active galactic nucleus (AGN) \citep{Moorwood1986, Roche1991, Weedman2005, Smith2007, AlonsoHerrero2016, Kirkpatrick2017, Jensen2017}, however with an on average smaller PAH equivalent. As such the equivalent width of the PAH features can be used to distinguish between AGN and/or strong starbursting galaxies from  normal, star-formation dominated systems \citep{Laurent2000, Brandl2006,  Sajina2007, Spoon2007, Pope2008b, Shipley2013, Esquej2014}. For AGN-dominated galaxies, the total IR emission may also arise from dust heated by the AGN rather than star formation activity, especially in wavelengths shorter than the peak of the FIR SED \citep[e.g..][]{Smith2007, Wu2010, Shipley2013, Mullaney2013}. Lower \lpahall/\lir\ ratios have previously been observed in AGN-dominated sources with respect to SFGs \citep{Armus2007, Sajina2008, Valiante2007} suggesting that PAHs at 6.2, 7.7, and 8.6 \si{\micro\metre} are suppressed due the presence of an AGN \citep{DiamondStanic2010}. Interestingly though, recent works in the local Universe report strong PAH 11.3 \si{\micro\metre} emission from the nuclear regions of Seyfert galaxies and QSOs \citep{Honig2010, AlonsoHerrero2014, Esquej2014, AlonsoHerrero2016}, indicating that PAH molecules could be excited, rather than destroyed, by the AGN itself \citep{Jensen2017}. 

A similar trend to \lpahall--\lir\ has been observed between the luminosity of the CO(1--0) transition line  (\lco), a common tracer of molecular gas, and the \lir. The majority of SFGs follow a tight relation between the SFR (traced by \lir) and the cold molecular gas (traced by \lco) or the total gas content (\mgas), which is known as the Kennicutt--Schmidt (KS) law spanning a large dynamical range \citep{Schmidt1959, Kennicutt1998}. 
Similar to the \lir--\lpah\ relation starbursting systems also appear as outliers in \lir--\lco\ relation, exhibiting an enhanced star formation efficiency (SFE=\lir/\lco) possibly driven by a major merger event, as supported by observations of local ULIRGs and a fraction of submillimeter galaxies (SMGs) at high redshift \citep{Rigopoulou1999, Pope2013}. 
The weaker PAH and CO emission (for a fixed \lir) in these star formation dominated galaxies can be explained by compact star forming regions and high SFEs due to a larger fraction of dense molecular gas \citep{Tacconi2008, Daddi2010a, Daddi2010b, DiazSantos2011, Pope2013, Kirkpatrick2014}. 
To this direction, \citet{Elbaz2011} found that the IR8=\lir/\lmid\footnote{\lmid\ is the monochromatic luminosity at rest-frame 8 \si{\micro\metre} as traced by the IRAC 8.0 \si{\micro\metre} band, which covers both the PAH 6.2, 7.7, and 8.6 \si{\micro\metre} complex at low redshift.} ratio can be used to separate normal SFGs with extended star-formation activity that also lay predominantly on the so-called "main-sequence" (MS) of galaxies \citep{Schreiber2015} from compact starbursts (SBs). Also, for star-formation dominated galaxies, \citet{Magdis2013} reported that IR8 variations are driven mainly by the strength of the PAH features rather than continuum variations, again indicative of more compact star formation for sources with weaker PAH features. 

Finally, several studies have revealed a connection between PAHs and the molecular gas (\mol) as traced by CO emission. Analyses of the observed radial profiles of PAH and CO emission in local galaxies indicate that PAHs can be used as a proxy for the molecular ISM in galaxies \citep{Regan2006}. The link between PAHs and molecular gas is further supported not only by the observed correlations between PAHs and CO emission on galaxy integrated scales \citep{Pope2013}, but also between between PAHs and cold dust emission at $ \geqslant 160$ \si{\micro\metre} \citep[e.g..][]{Haas2002, Bendo2008, Jones2015}.

In this work, we further explore the connection between the PAHs and the molecular gas of galaxies, with new single-dish CO(1--0) line observations of 34 IR-bright PAH emitting SFGs across the MS selected from the 5MUSES survey \citep{Wu2010}, increasing the existing sample of PAH, IR, and CO detected galaxies at intermediate redshifts ($0.03 < z < 0.28$) by a factor of 2.4  (Section \ref{sec:data}). We complement our sample with existing CO(1--0) and PAH observations from the literature in order to determine the scatter of the scaling relations between IR, PAH, and CO data spanning 2 orders of magnitude in luminosity and covering a broad range of redshifts ($0<z<4$). In Section \ref{sec:pah_ir} we present the \lpah--\lir\ for normal SFGs and identify local ULIRGs and high-$z$ SBs as clear outliers characterised by lower \lpah/\lir\ ratios. In section \ref{sec:co_ir} we show that these outliers also exhibit lower \lco/\lir\ ratios compared to the \lco--\lir\ relation defined by the general population of normal galaxies. On the other hand, in Section \ref{sec:co_pah} we present a universal \lpah--\lco\ relation followed by both normal SFGs and SBs at all redshifts. This, along with the strong correlation between the PAH and cold dust emission ($\lambda \geqslant 160$ \si{\micro\metre}) presented in Section \ref{sec:pah_dust},  
motivates us to explore PAHs as a proxy for the molecular gas in Section \ref{sec:mgas}. 

Throughout this paper we adopt a standard cosmology with $H_{0} = 70$ \si{\km\per\s\per\mpc}, $\Omega_{\rm M} = 0.30$, and $\Omega_{\Lambda} = 0.70$.

\section{Data sample}\label{sec:data}
We have selected 34 star-forming targets from the 5 mJy Unbiased \s\ Extragalactic Survey \citep[5MUSES:][]{Wu2010}, in order to examine the gas and ISM properties of star-formation dominated galaxies at intermediate redshift ($0.03<z<0.28$) by detecting CO(1--0) emission and using existing observations. 5MUSES is a 24 \si{\micro\metre} flux-limited ($f_{24\mu {\rm m}} > 5$ mJy) spectroscopic survey with \s\ IRS, containing 330 galaxies with \lir$\sim 10^{10}-10^{12}$ \lsol\ located in the SWIRE \citep{Lonsdale2003} and Extragalactic First Look Survey (XFLS) fields \citep{Fadda2006}. The sample fills out the gap between local SFGs \citep{Kennicutt2003, Smith2007, Dale2009}, low-$z$ ULIRGs \citep{Armus2007, Desai2007, Veilleux2009}, and more distant galaxies with available spectroscopy data \citep{Houck2005, Yan2007}. In addition, the full sample has \s\ Infrared Array Camera observations \citep[IRAC:][]{Fazio2004} at $3.6-8$ \si{\micro\metre} and Multiband Imaging Photometer \citep[MIPS:][]{Rieke2004} at $70-160$ \si{\micro\metre} \citep{Wu2010}, where 90\% and 54\% of the galaxies are detected at 70 and 160 \si{\micro\metre}, respectively.

From the 5MUSES sample, 280 galaxies have spectroscopically confirmed redshifts with low-resolution ($R=64-128$) MIR spectra which were collected using the short-low (SL: 5.5--14.5 \si{\micro\metre}) and long-low (LL: 14--35 \si{\micro\metre}) spectral modules of the \s\ InfraRed Spectrograph \citep[IRS]{Houck2004} as described in \citet{Wu2010}. In addition to \s\ IRAC, MIPS, and IRS observations, a subsample of 188 galaxies  (with spectroscopically confirmed redshifts) have FIR photometric coverage at 250, 350, and 500 \si{\micro\metre} obtained with the \textit{Herschel Space Observatory} \citep{Griffin2010} through SPIRE observations as part of the \h\ Multi-Tiered Extragalactic Survey \citep[HerMES:][]{Oliver2010, Oliver2012}. Out of the 188 sources, a flux density limit of $S_{\rm \nu} > 15$ mJy in the \h\ SPIRE bands yields a detection for 154 (82\%), 108 (57\%), and 50 (27\%) sources at 250 \si{\micro\metre}, 350 \si{\micro\metre} and 500 \si{\micro\metre}, respectively \citep{Magdis2013}. Stellar masses of the full 5MUSES sample have been estimated by \citet{Shi2011} using the \citet{BruzualCharlot2003} population synthesis model to fit optical and IR photometry assuming a \citet{Chabrier2003} IMF.

\begin{figure}
	\centering
	\includegraphics[width=0.47\textwidth, trim={0 0 0 0}, clip]{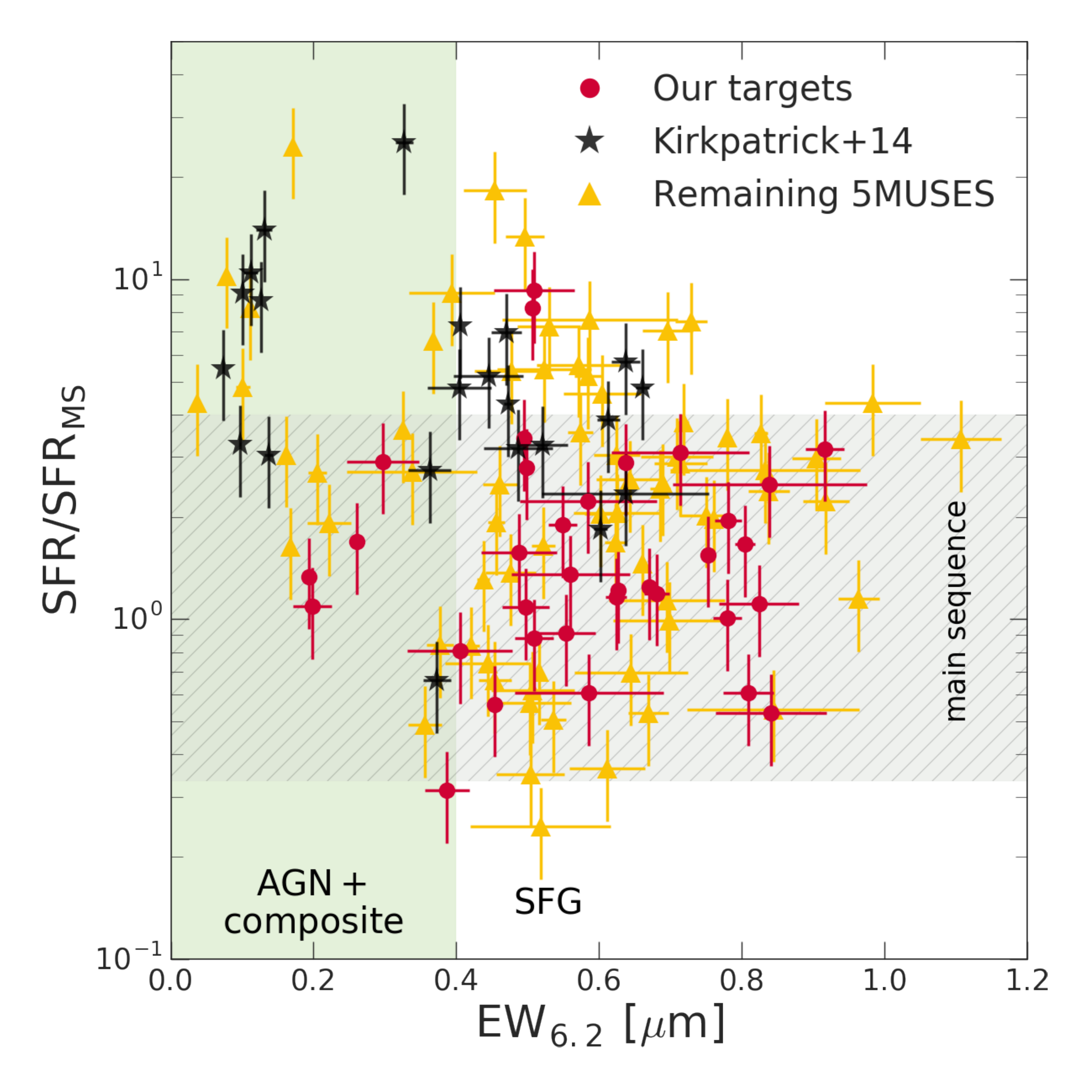}
    \caption{The equivalent width of the PAH 6.2 \si{\micro\metre} feature (\ew) 
    versus offset from the MS (grey shaded region). Red points 
    depict galaxies selected for CO(1--0) line observations presented in this study. 
    Previously CO-detected 5MUSES galaxies from \protect\citet{Kirkpatrick2014} 
    are shown in black, whereas the rest of the 5MUSES sources are shown in yellow. 
    Galaxies with \ew$\leq0.4$ \si{\micro\metre} are classified as AGN-dominated or composite sources 
    (green region).}
    \label{fig:ew_dist_ms}
\end{figure}

In Figure \ref{fig:ew_dist_ms}, we present the equivalent width of the PAH 6.2 \si{\micro\metre} (\ew) vs. distance from the MS for the full 5MUSES sample. The offset from the MS, ${\rm SFR}/{\rm SFR}_{\rm MS}$($z$, \Mstar), is determined by adopting eq. 9 in \citet{Schreiber2015} after converting our stellar masses from a \citet{Chabrier2003} IMF to a \citet{Salpeter1955} IMF using $M_{\ast}^{\rm S} = 1.70 \times M_{\ast}^{\rm C}$ \citep{Speagle2014}. SFRs are derived using \lir\ estimates (Section \ref{sec:sed}). 
We use optical spectroscopy and/or the \ew\ to identify AGN-dominated sources in the sample. In absence of optical spectroscopy, we classify sources with \ew$\leq 0.4$ \si{\micro\metre} as AGN and composite sources \citep[see][for a detailed AGN characterisation of the 5MUSES sample]{Wu2010, Magdis2013}. Since we aim at examining the ISM properties of normal galaxies at intermediate redshifts, we primarily selected targets with \ew$> 0.4$ \si{\micro\metre} across the MS (${\rm SFR}/{\rm SFR}_{\rm MS} < 4$) for follow-up CO(1--0) line observations. Moreover, all of our targets have \s\ and \h\ observations, \lir$=10^{9.2}-10^{11.8}$ \lsol, and stellar masses of $\langle$\ms$\rangle = 10^{10}$~\msol. 

 \begin{figure*}
 \begin{subfigure}{0.3\textwidth}
	 \includegraphics[width=\linewidth] {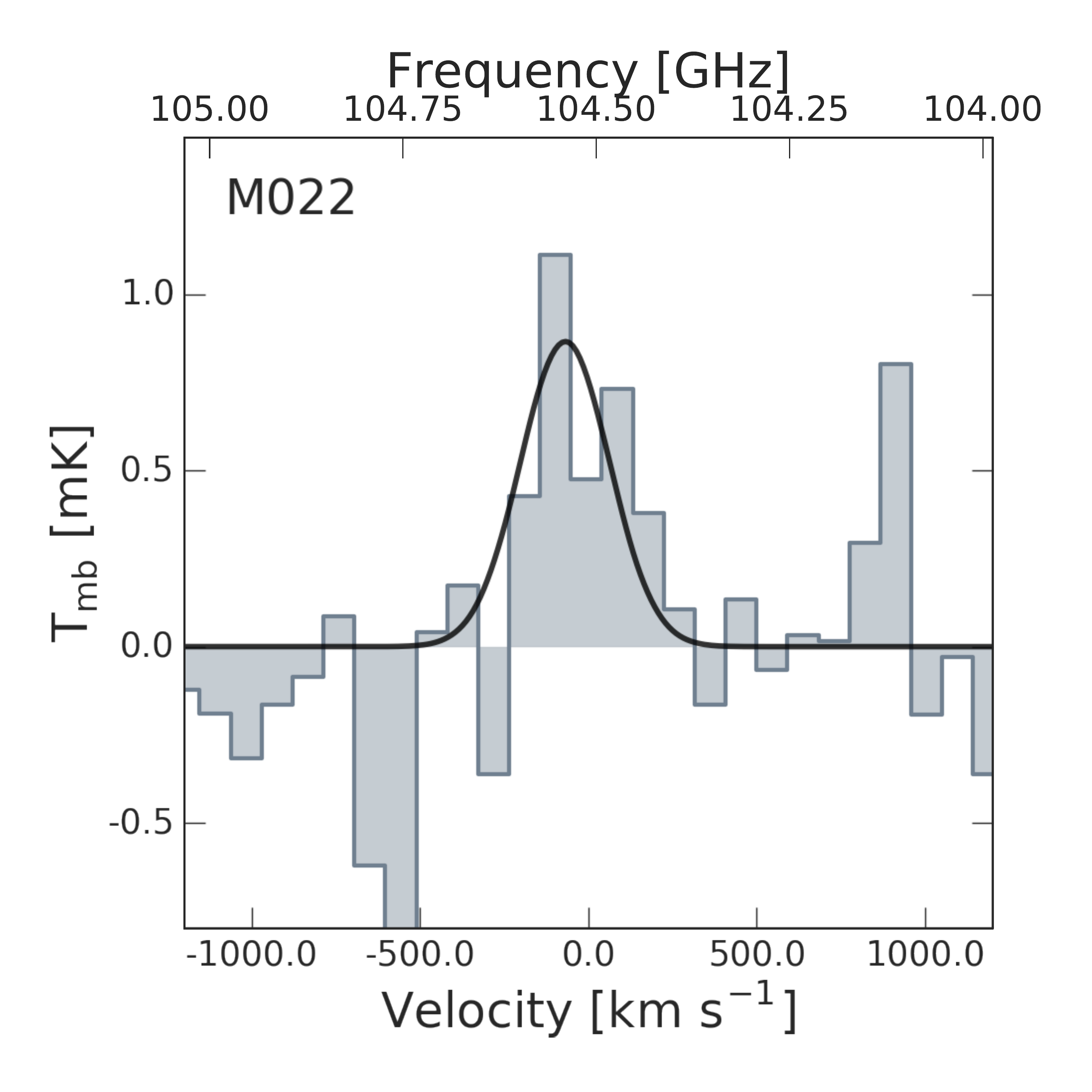}
 \end{subfigure}
 \begin{subfigure}{0.3\textwidth}
	 \includegraphics[width=\linewidth]{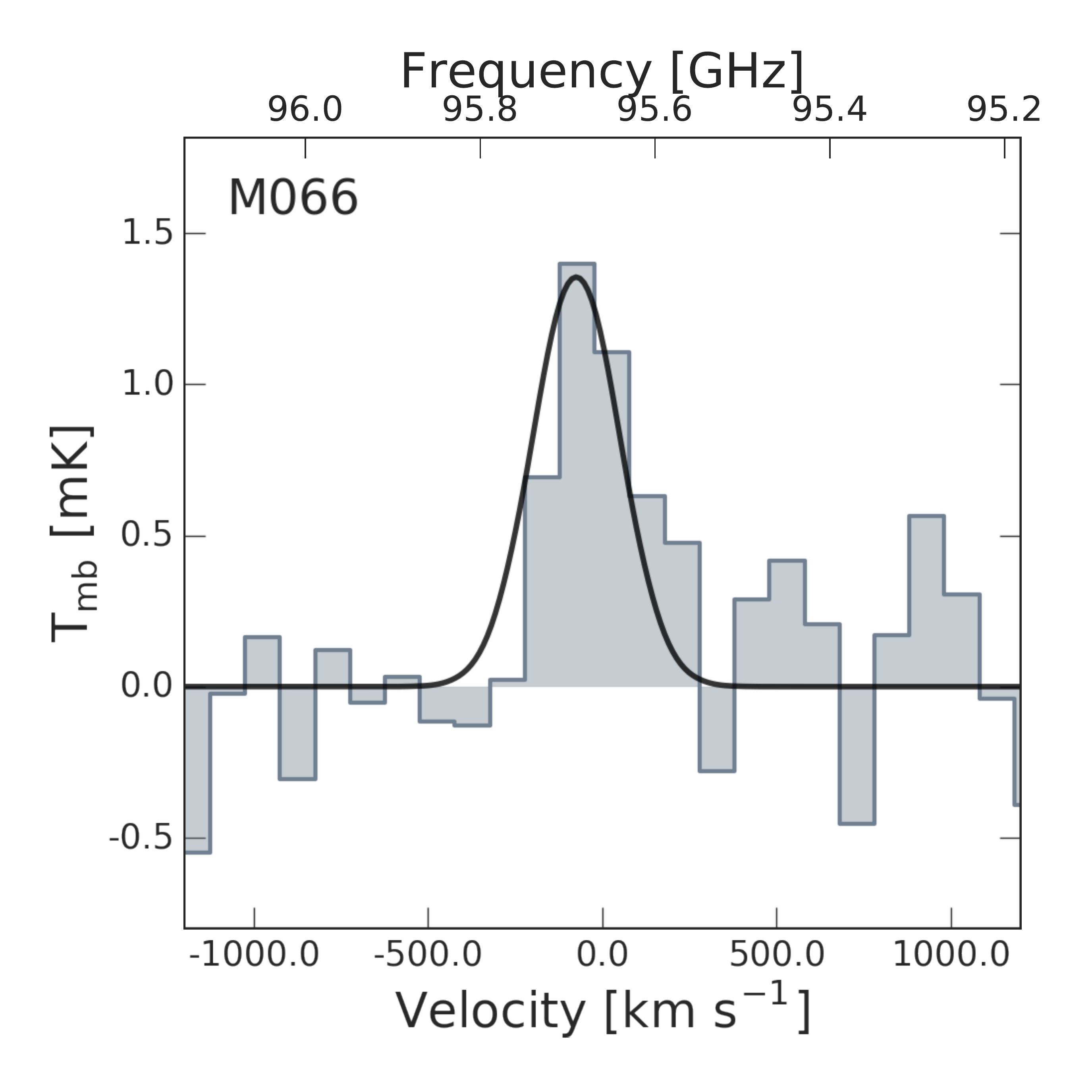}
 \end{subfigure}
 \begin{subfigure}{0.3\textwidth}
	 \includegraphics[width=\linewidth]{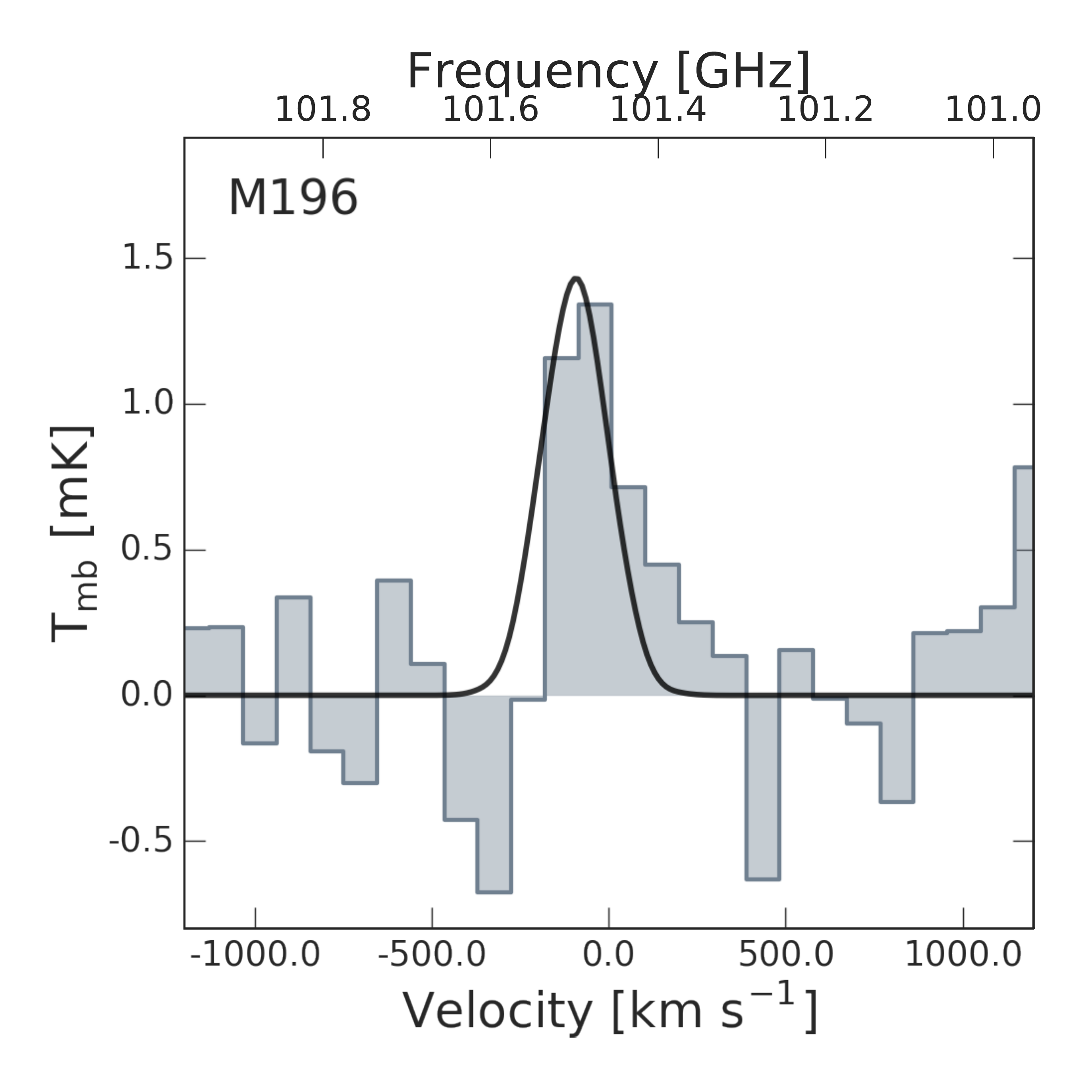}
 \end{subfigure}
 
 \medskip
 \begin{subfigure}{0.3\textwidth}
	 \includegraphics[width=\linewidth]{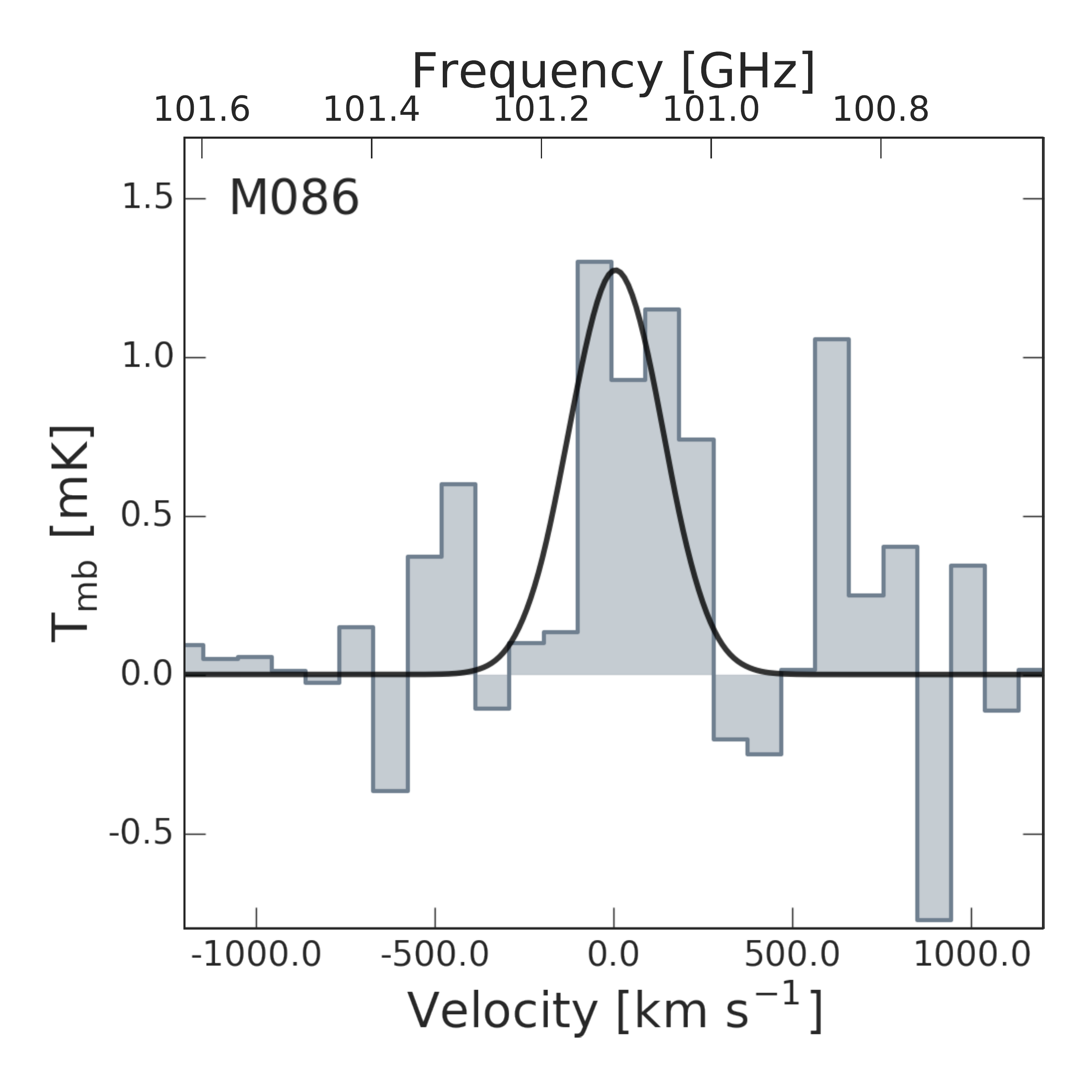}
 \end{subfigure}
 \begin{subfigure}{0.3\textwidth}
	 \includegraphics[width=\linewidth]{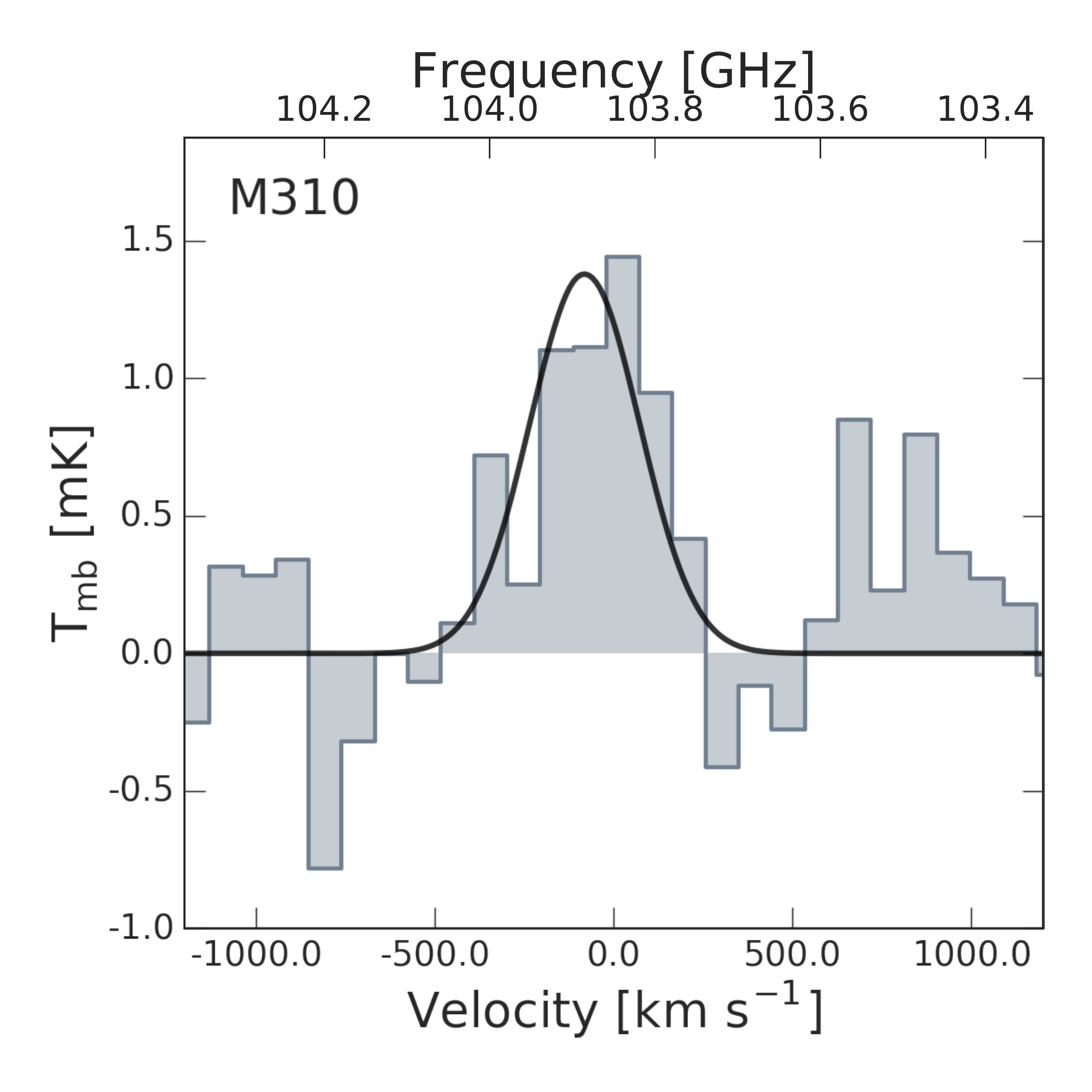}
 \end{subfigure}
 \begin{subfigure}{0.3\textwidth}
	 \includegraphics[width=\linewidth]{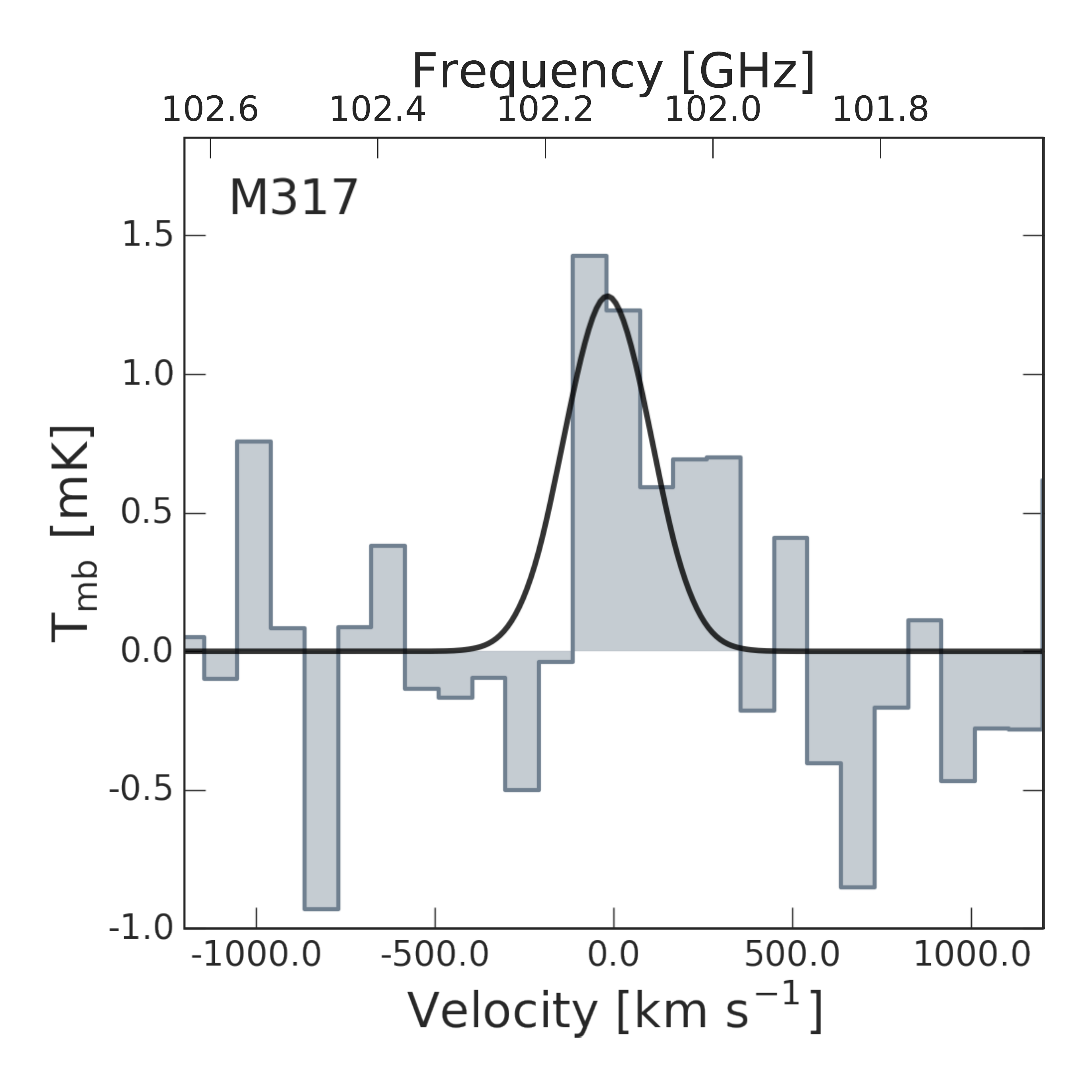}
 \end{subfigure}

\medskip
\begin{subfigure}{0.3\textwidth}
	 \includegraphics[width=\linewidth] {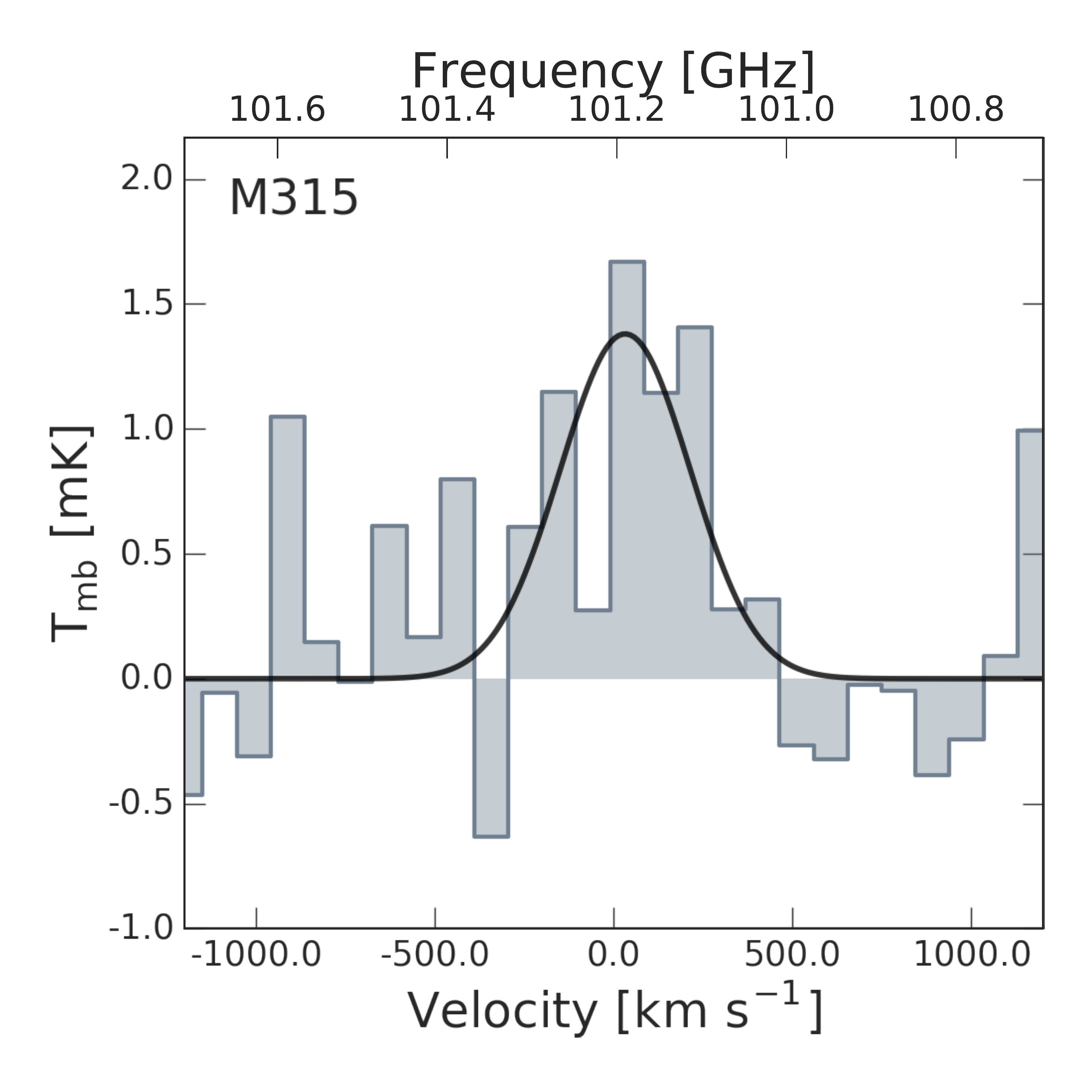}
 \end{subfigure}
 \begin{subfigure}{0.3\textwidth}
	 \includegraphics[width=\linewidth]{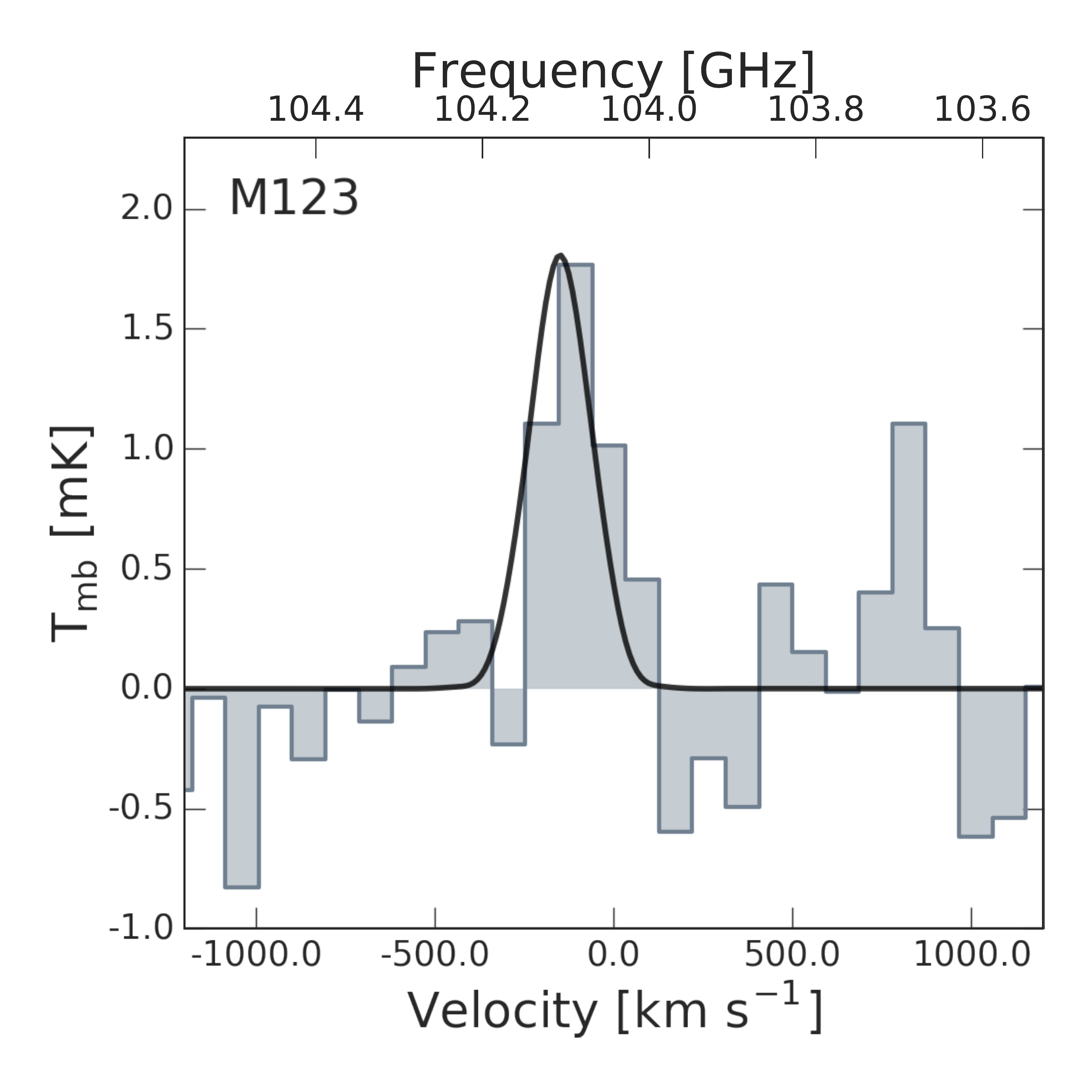}
 \end{subfigure}
 \begin{subfigure}{0.3\textwidth}
	 \includegraphics[width=\linewidth]{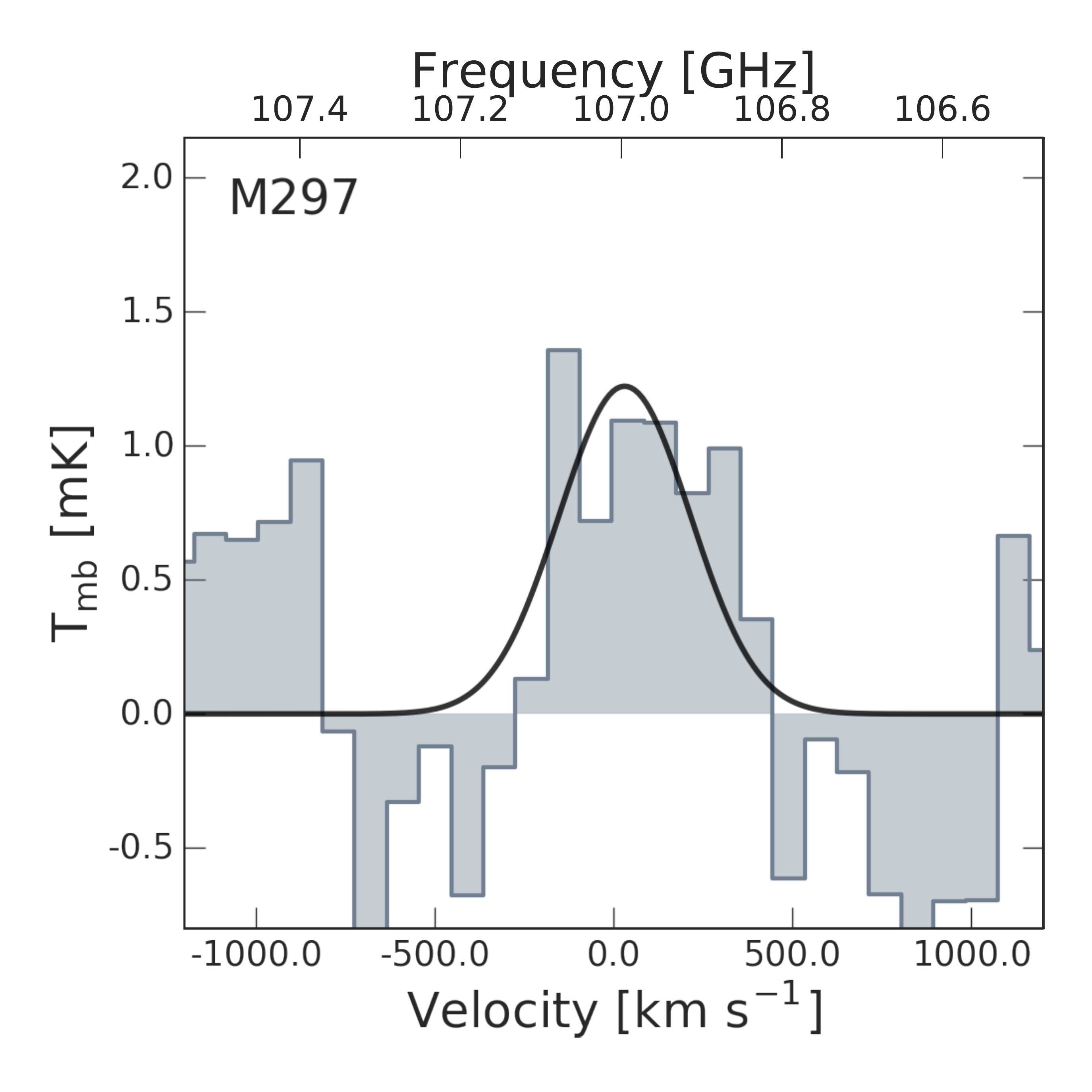}
 \end{subfigure}
	 \caption{CO(1--0) spectra with antenna temperature [mK] as a function of velocity [\si{\km\per\s}] of CO-detected 5MUSES galaxies followed-up with IRAM/EMIR. All spectra are smoothed to a  velocity resolution of $\sim 70$ \si{\km\per\s}. A line is considered detected if the integrated signal is above $3\sigma$. The black line shows the best-fit Gaussian profile to the observed CO line.} 
	 \label{fig:subplots_co1}
 \end{figure*}
 
\subsection{New CO(1--0) line observations}\label{sec:co_obs}
The single-dish observations were carried out with the IRAM 30 m telescope at Pico Veleta, Spain, in July 2015 and June and September 2016. All galaxies were observed at 3 mm using the spectral line receiver band E0 of EMIR with WILMA as backends in order to observe the CO(1--0) emission line. The receiver was tuned to the expected frequency of the targets (in the range 95 GHz < $\nu$ < 107 GHz) and the wobbler switching mode was used. We spent $1-8$ hours on each galaxy. During observations the pointing of the telescope was checked every two hours using a bright nearby source. The velocity-integrated CO line intensities were converted from antenna temperature scale ($T_{\rm a}^{*}$) to Jy using $S/T_{\rm a}^{*} = 6.2$ \si{\Jy\per\K}. The CO(1--0) line luminosities were estimated in units of [\ulco] using the following equation from \citet{Solomon2005}: 
\begin{eqnarray} \label{eq:lco}
L^{\prime}_{\rm CO} = 3.25 \times 10^7 S_{\rm CO} \Delta v~ 
\nu_{\rm obs}^{-2}~ D_{\rm L}^{2} (1+z)^3
\end{eqnarray}
where $S_{\rm CO} \Delta v$ [\si{\Jy\km\per\s}] is the velocity integrated flux, $\nu_{\rm obs}$ [GHz] is the observed CO(1--0) frequency, and $D_{\rm L}$ [Mpc] is the luminosity distance. 

Spectra were reduced using the CLASS/GILDAS\footnote{\url{http://www.iram.fr/IRAMFR/GILDAS}} software, where each galaxy spectrum was averaged and smoothed to a velocity resolution of $70-100$ \si{\km\per\s}. Linear baselines were assumed for all targets. The spectra of galaxies with detected CO(1--0) line emission are shown in Figure \ref{fig:subplots_co1}. A detected CO line is considered when the integrated signal is above $3 \sigma$. We detect significant CO line emission in 24 galaxies, whereas upper limits are determined for the remaining targets assuming a CO(1--0) line width of 300 \si{\km\per\s}. The CO luminosities and the observed properties of each galaxy from our observing runs are listed in Table \ref{tab:muses_co}.

\begin{figure*}
\ContinuedFloat
 \begin{subfigure}{0.3\textwidth}
	 \includegraphics[width=\linewidth]{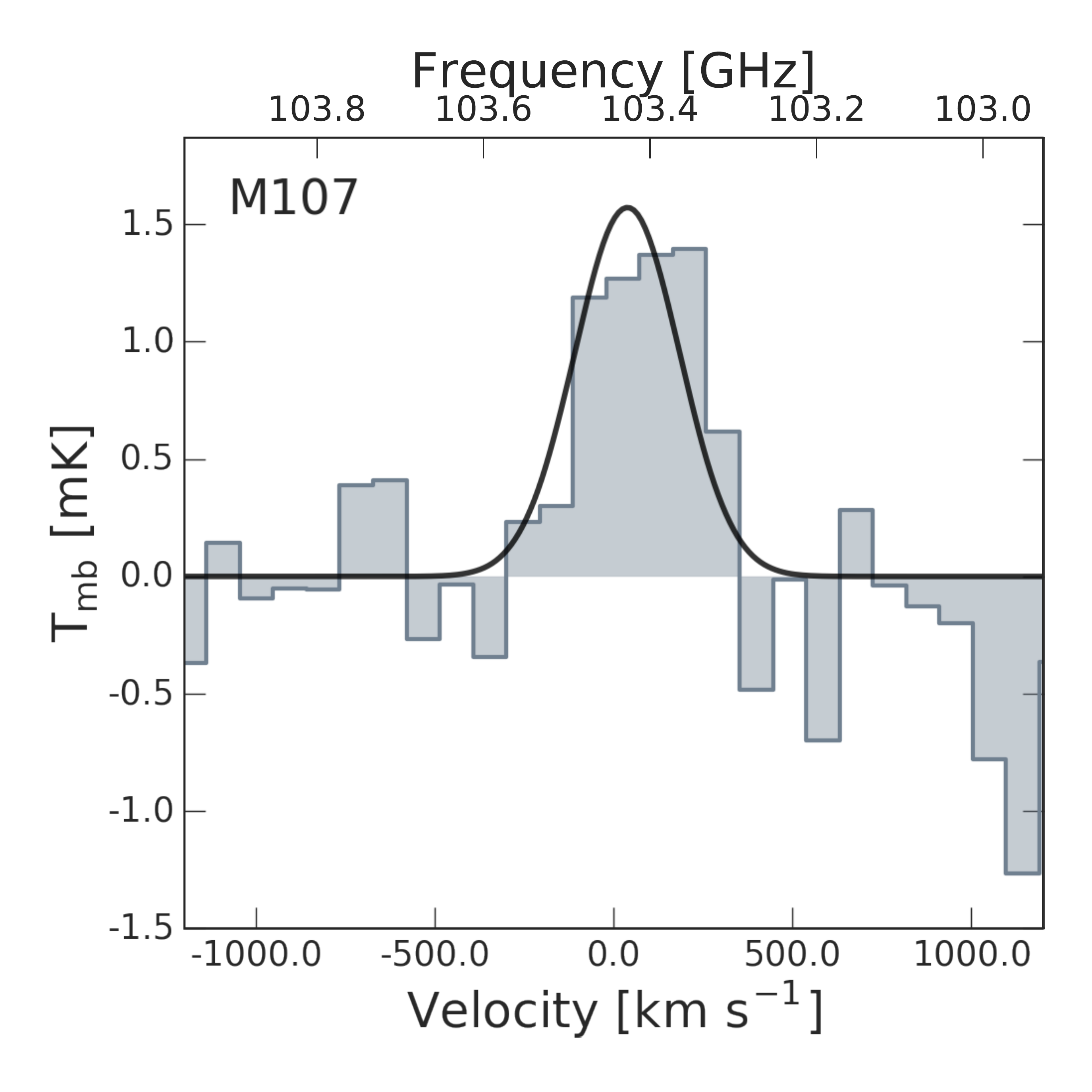}
 \end{subfigure}
 \begin{subfigure}{0.3\textwidth}
	 \includegraphics[width=\linewidth]{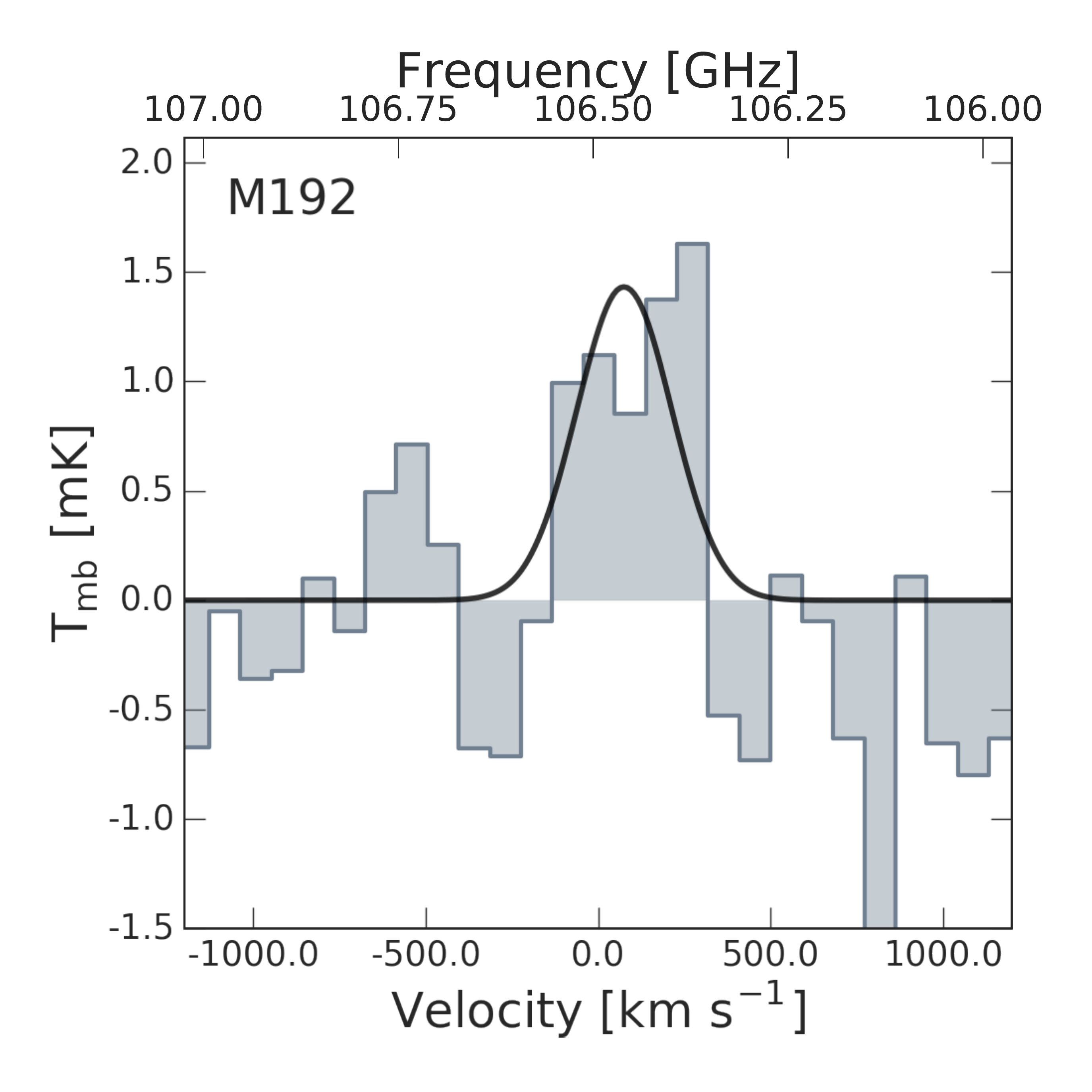}
 \end{subfigure}
 \begin{subfigure}{0.3\textwidth}
	 \includegraphics[width=\linewidth]{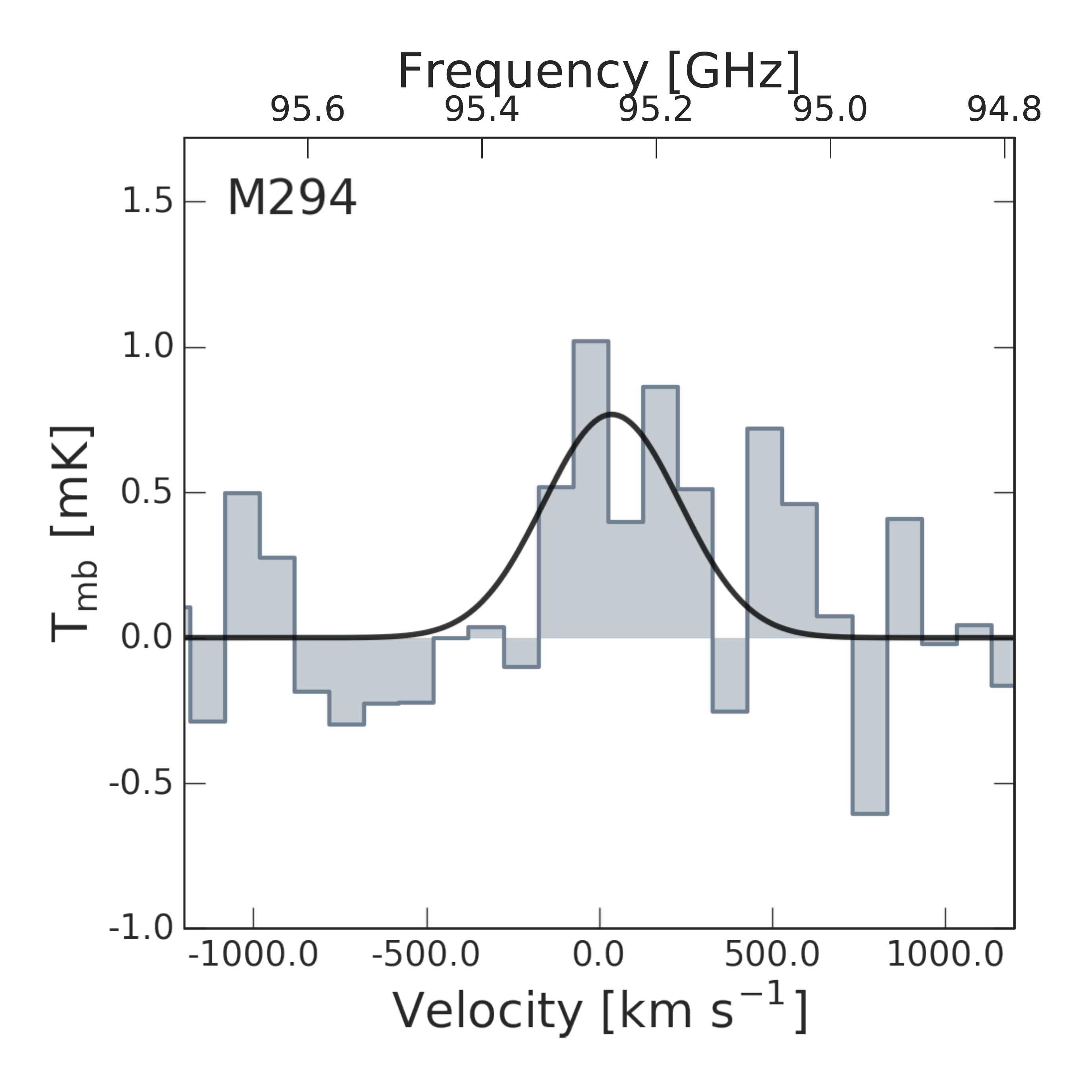}
 \end{subfigure}

\medskip
 \begin{subfigure}{0.3\textwidth}
	  \includegraphics[width=\linewidth] {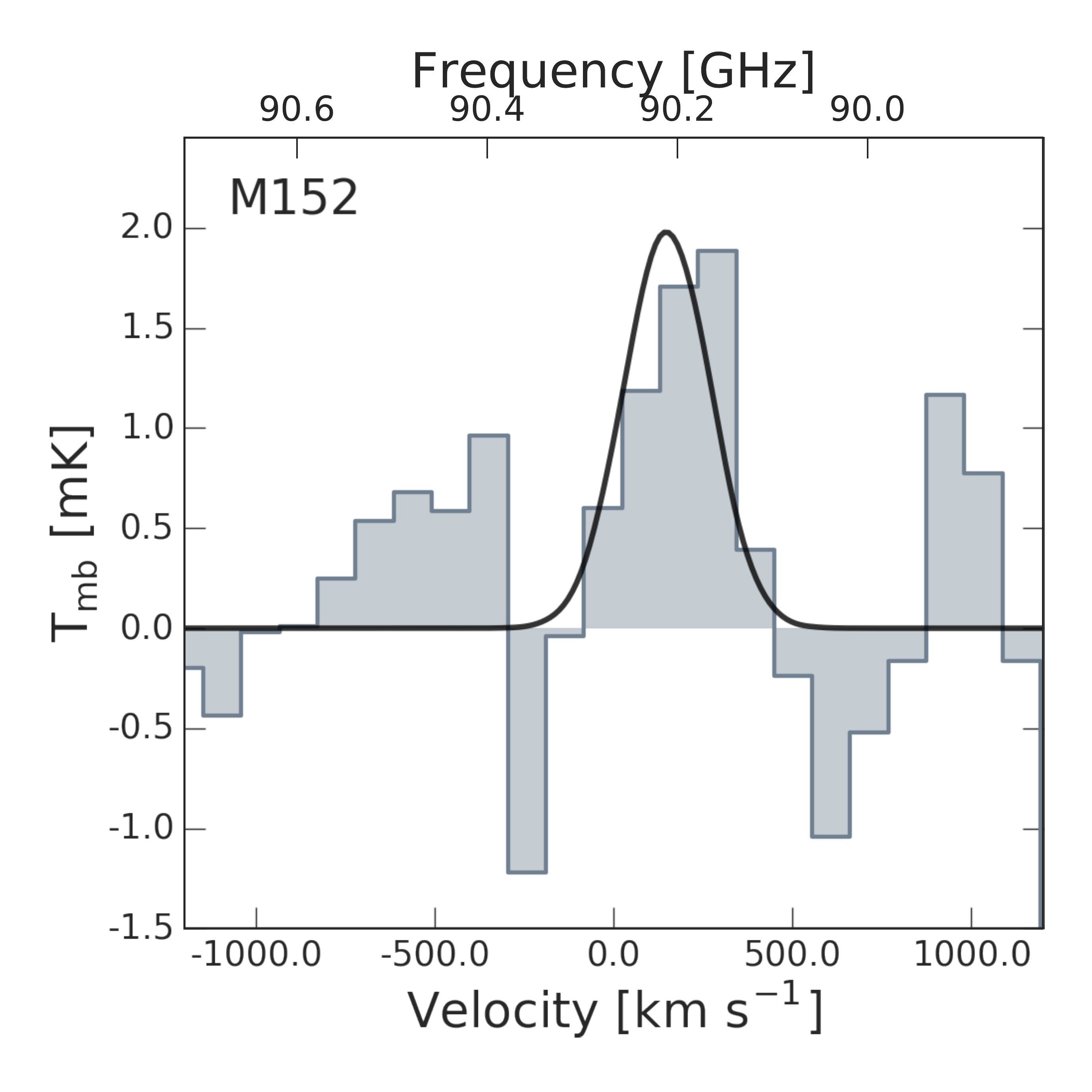}
  \end{subfigure}
  \begin{subfigure}{0.3\textwidth}
	  \includegraphics[width=\linewidth]{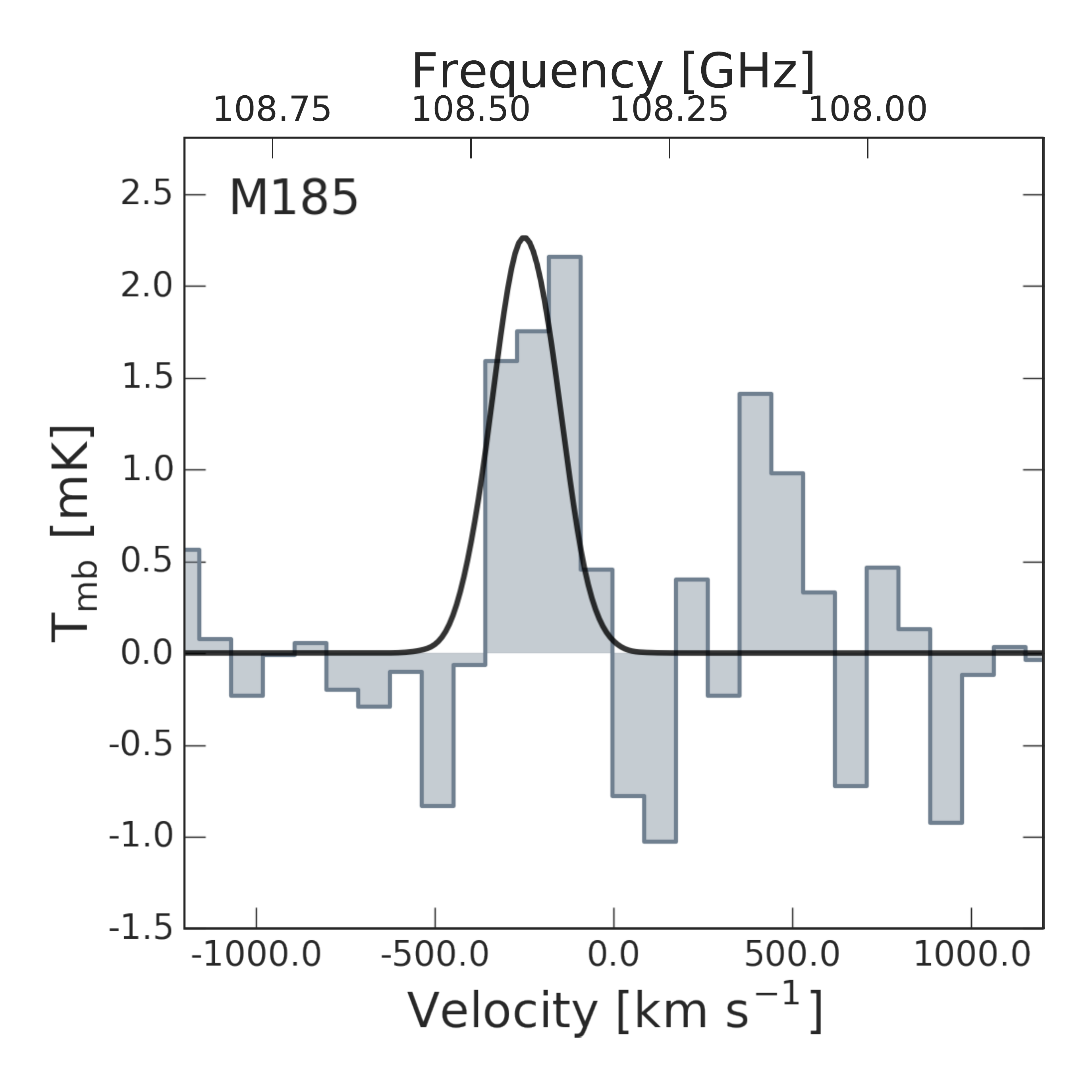}
  \end{subfigure}
  \begin{subfigure}{0.3\textwidth}
	  \includegraphics[width=\linewidth]{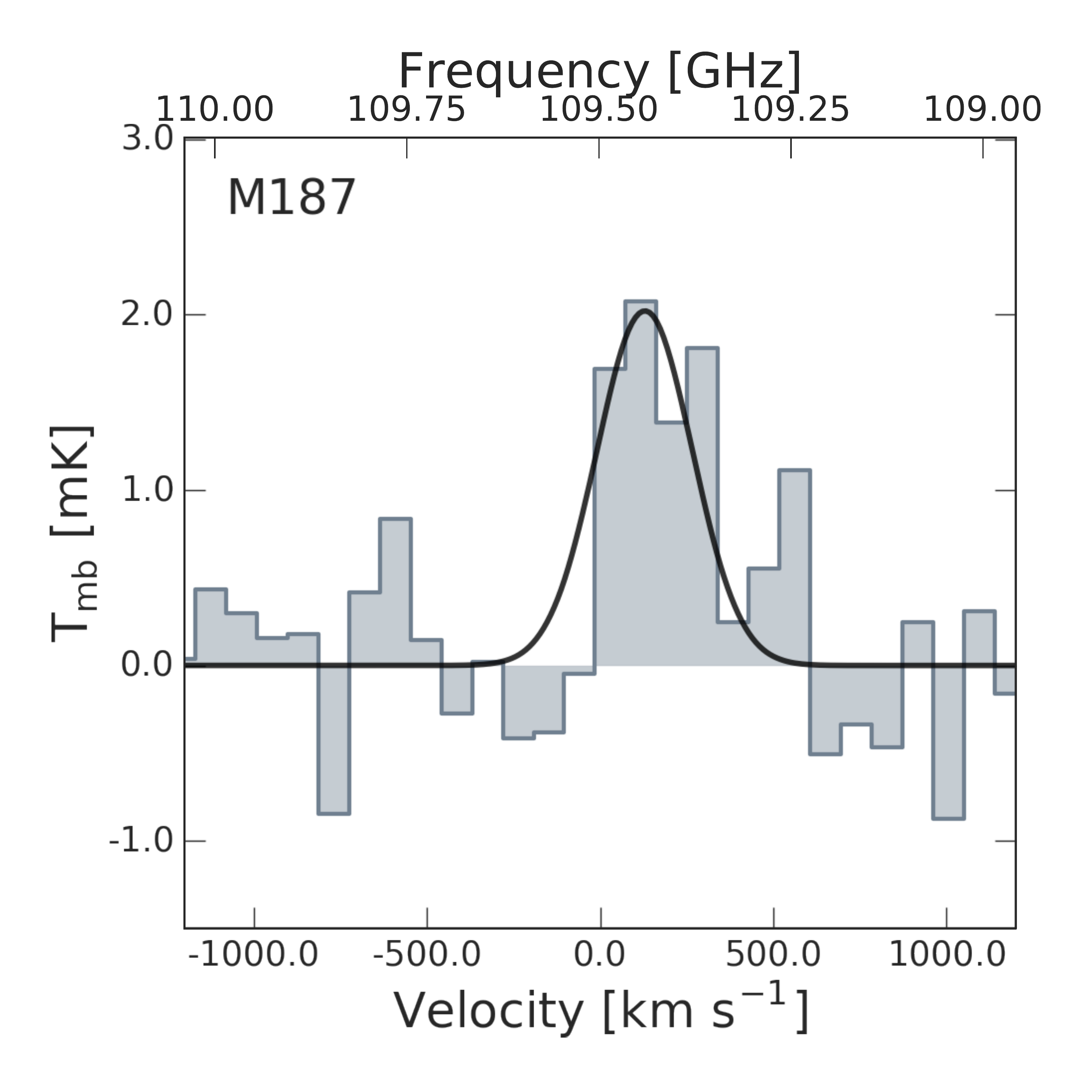}
  \end{subfigure}
  
  \medskip
  \begin{subfigure}{0.3\textwidth}
	  \includegraphics[width=\linewidth]{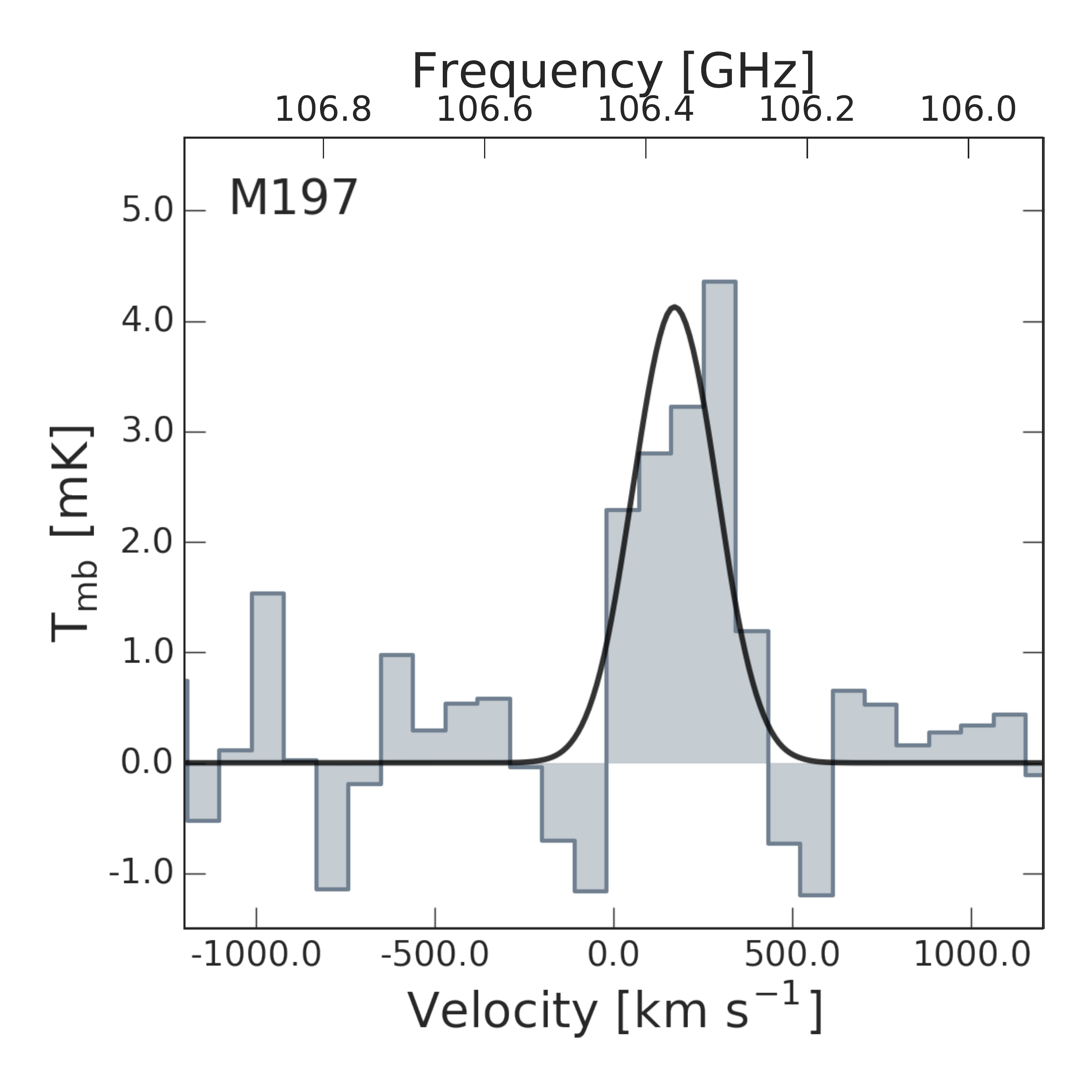}
  \end{subfigure}
  \begin{subfigure}{0.3\textwidth}
	  \includegraphics[width=\linewidth]{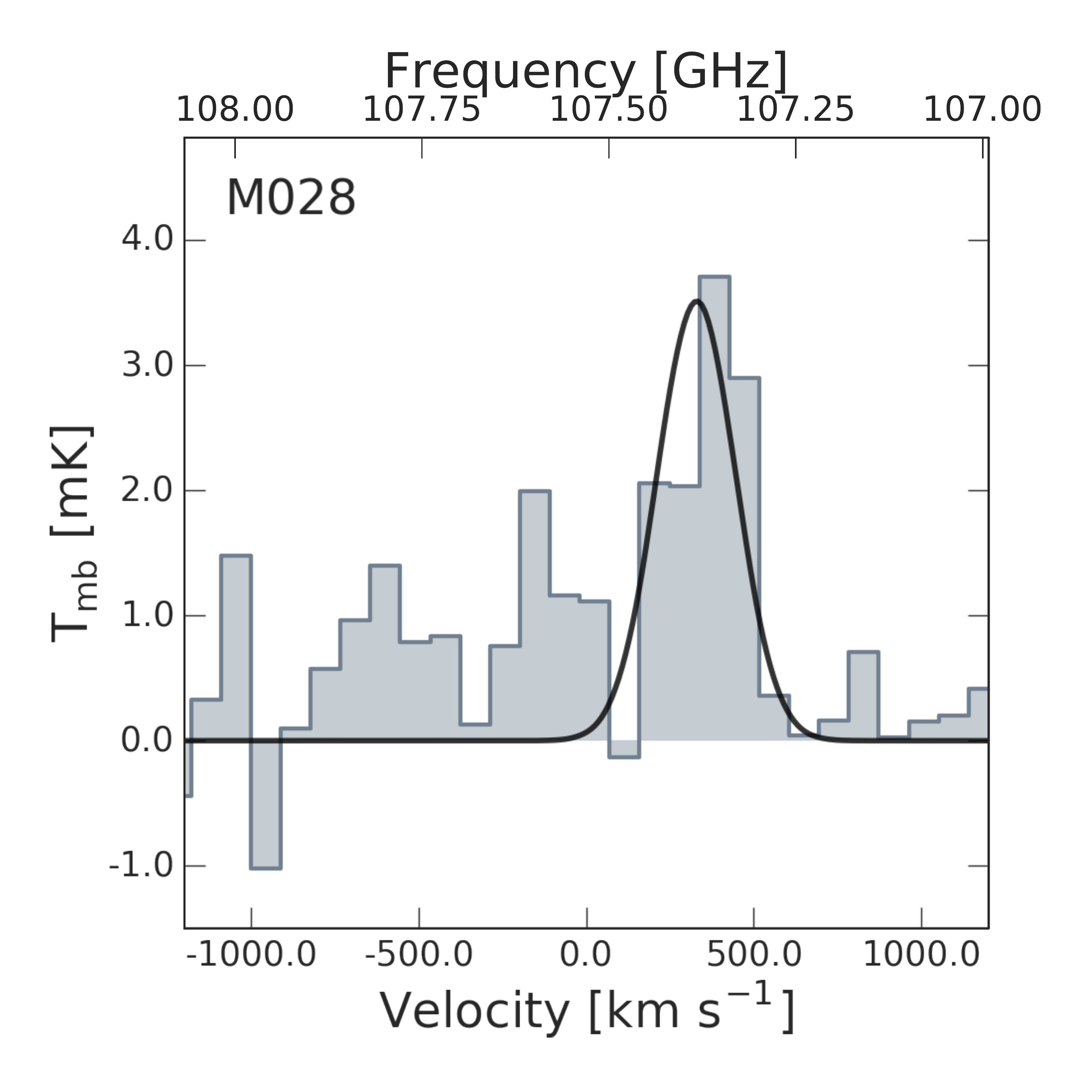}
  \end{subfigure}
  \begin{subfigure}{0.3\textwidth}
	  \includegraphics[width=\linewidth]{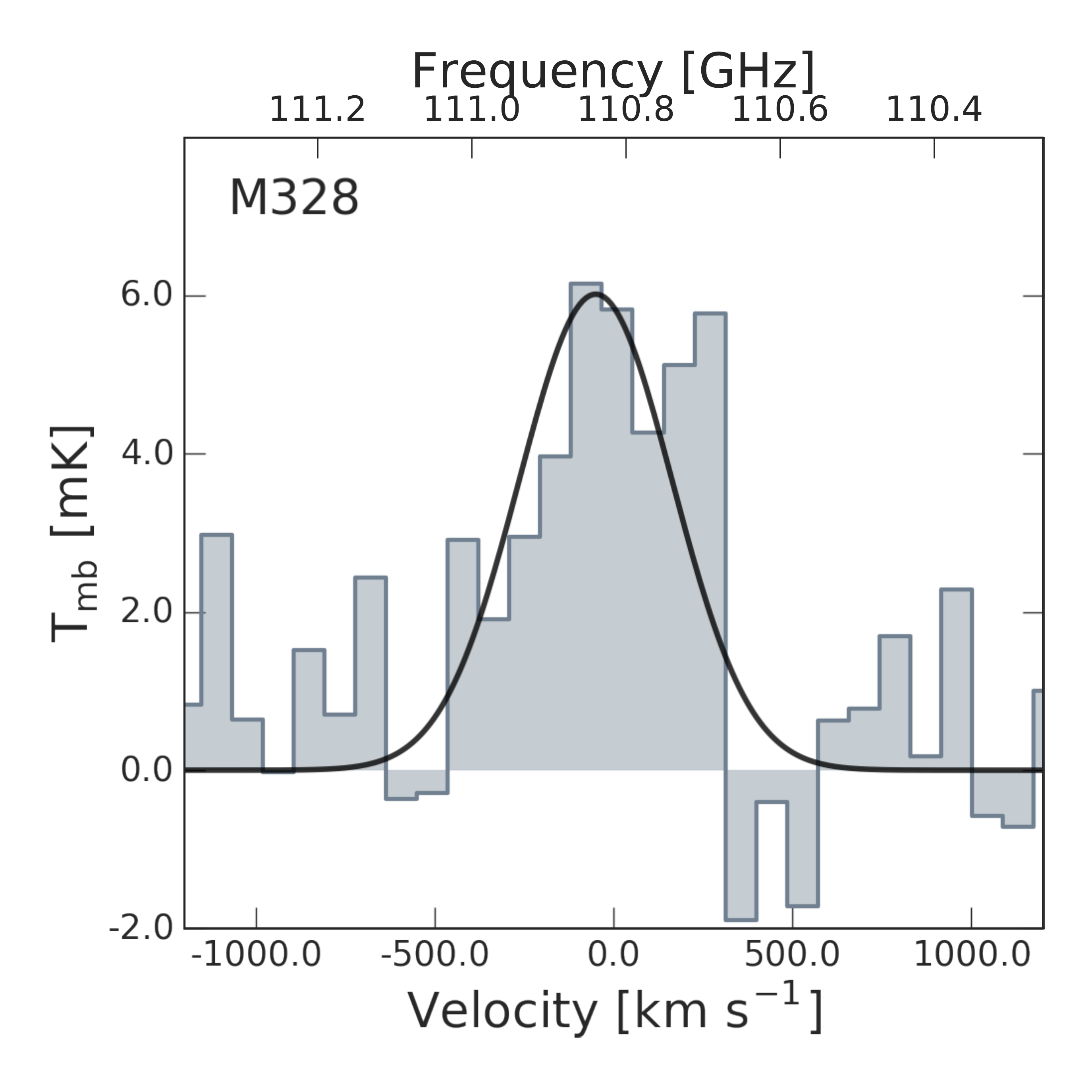}
  \end{subfigure}
	  \caption{(Continued)} 
\end{figure*}
 
\begin{figure*}
 \medskip
  \begin{subfigure}{0.3\textwidth}
 	 \includegraphics[width=\linewidth]{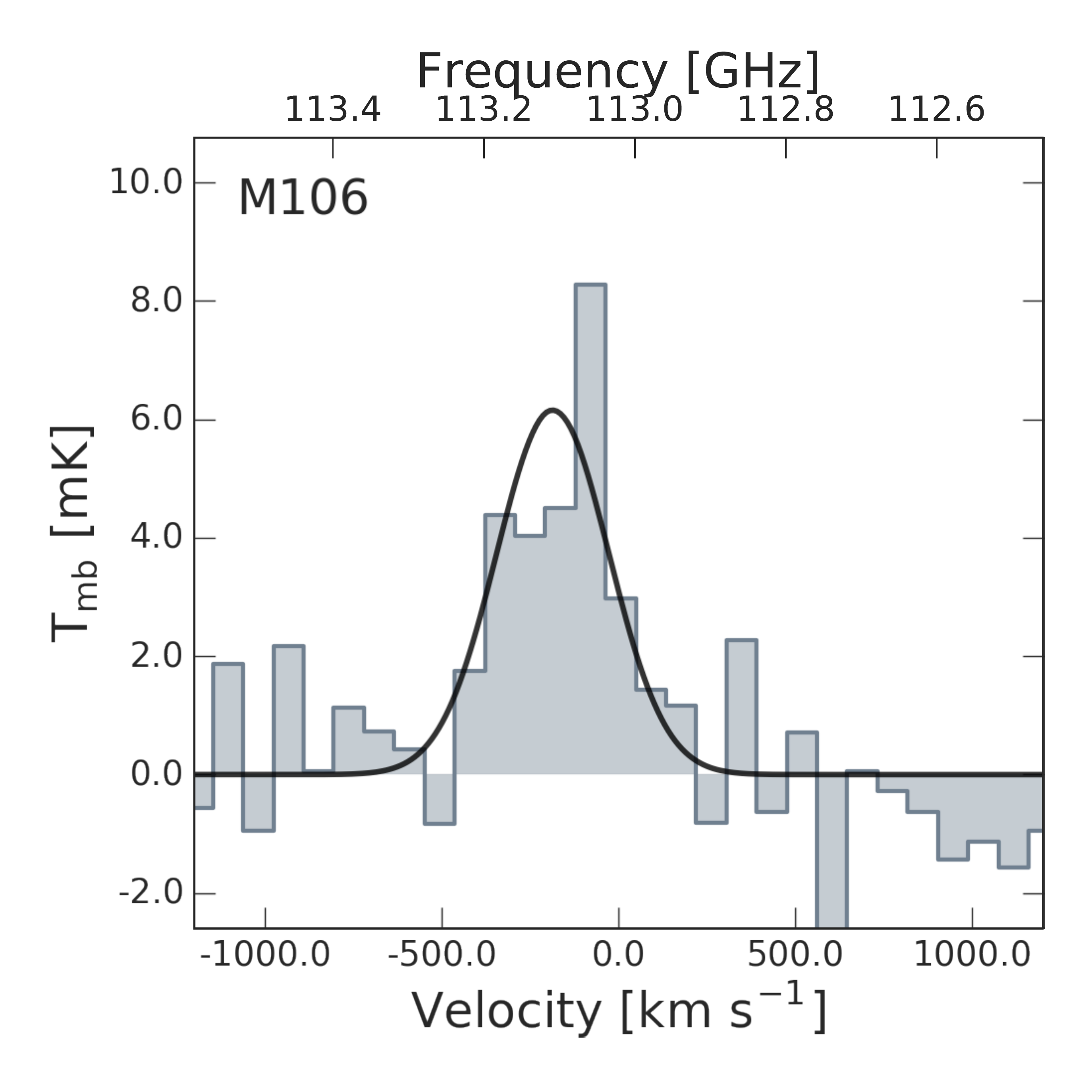}
  \end{subfigure}
  \begin{subfigure}{0.3\textwidth}
 	 \includegraphics[width=\linewidth]{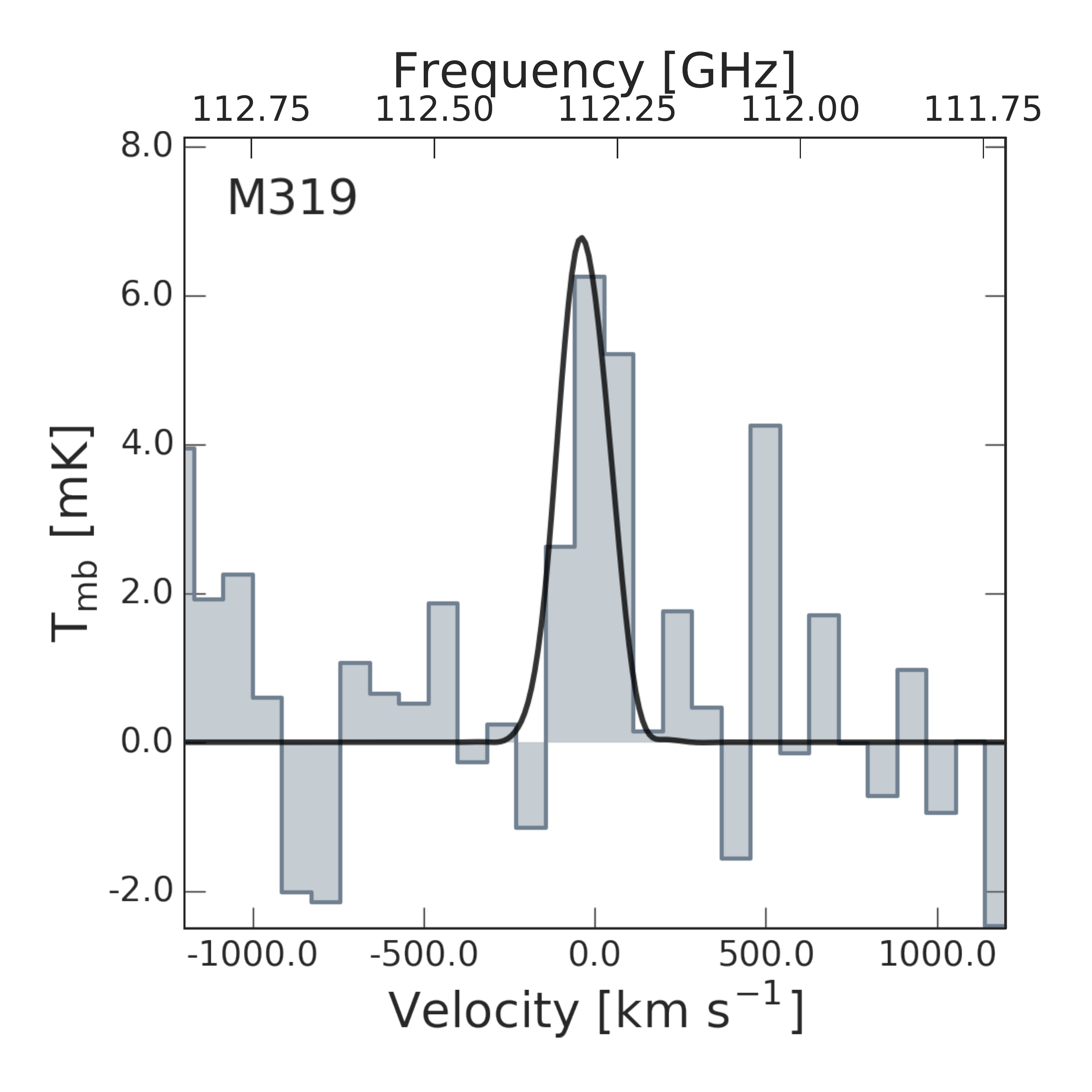}
  \end{subfigure}
  \begin{subfigure}{0.3\textwidth}
 	 \includegraphics[width=\linewidth]{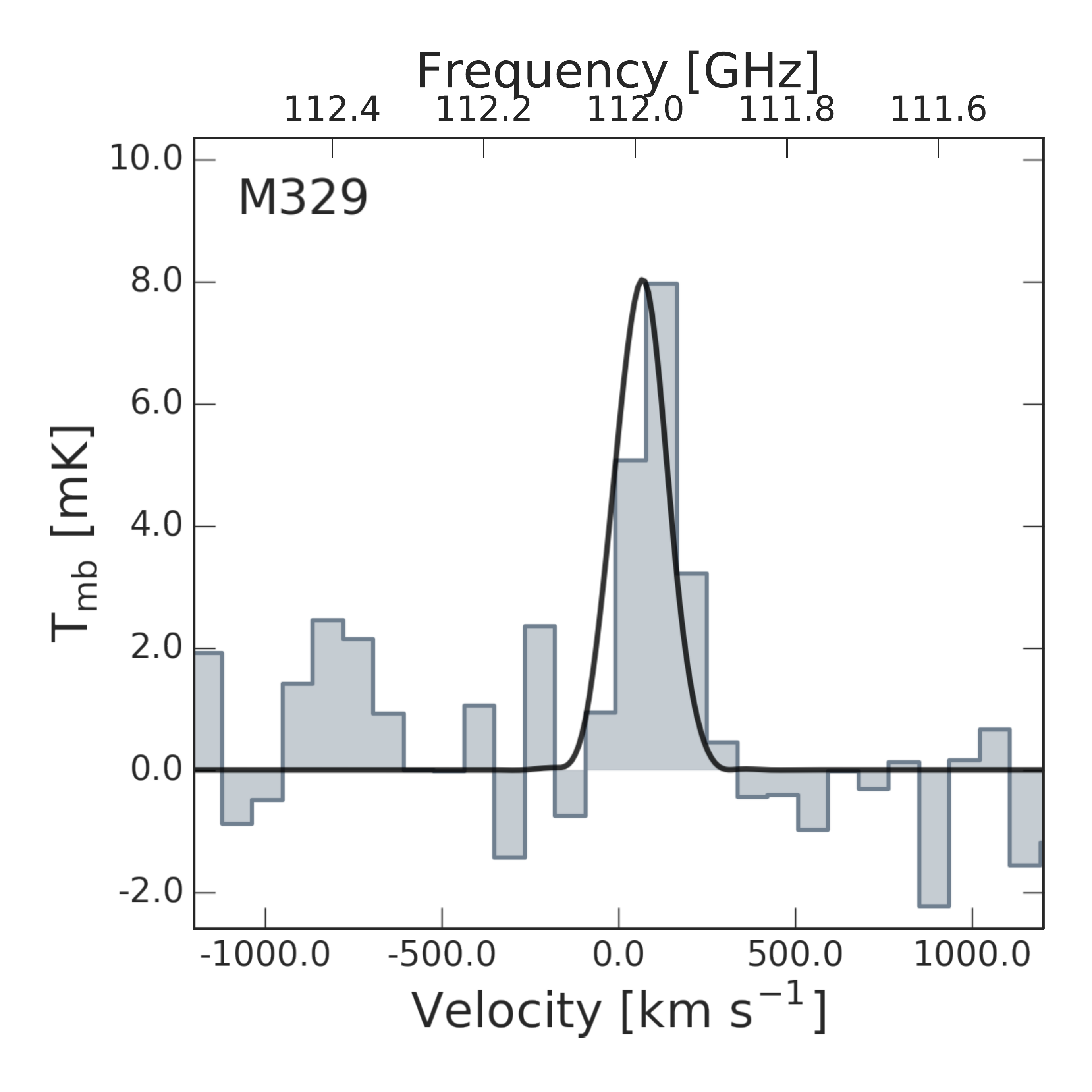}
  \end{subfigure}
  
	\medskip 
 \begin{subfigure}{0.3\textwidth}
	  \includegraphics[width=\linewidth] {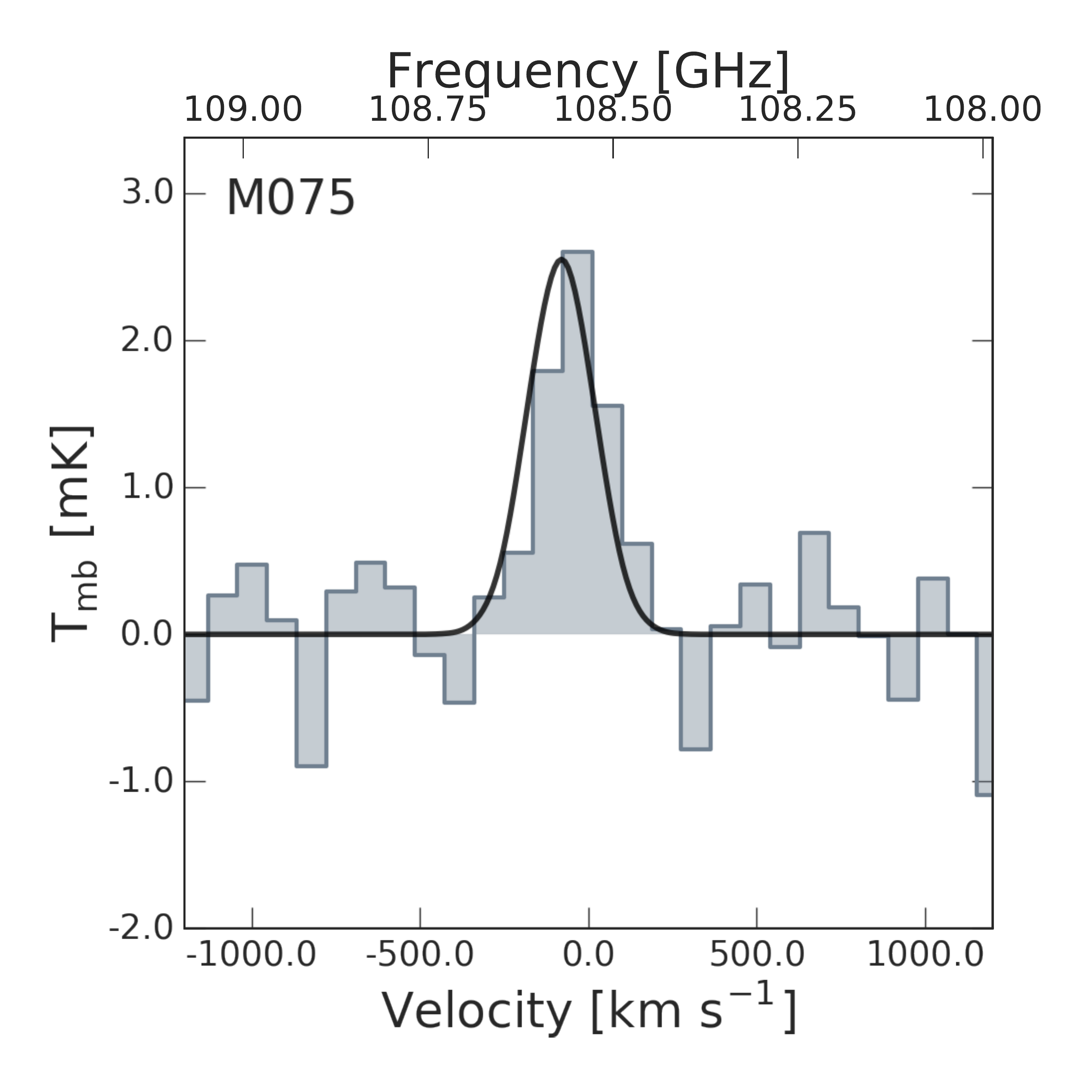}
  \end{subfigure}
  \begin{subfigure}{0.3\textwidth}
	  \includegraphics[width=\linewidth]{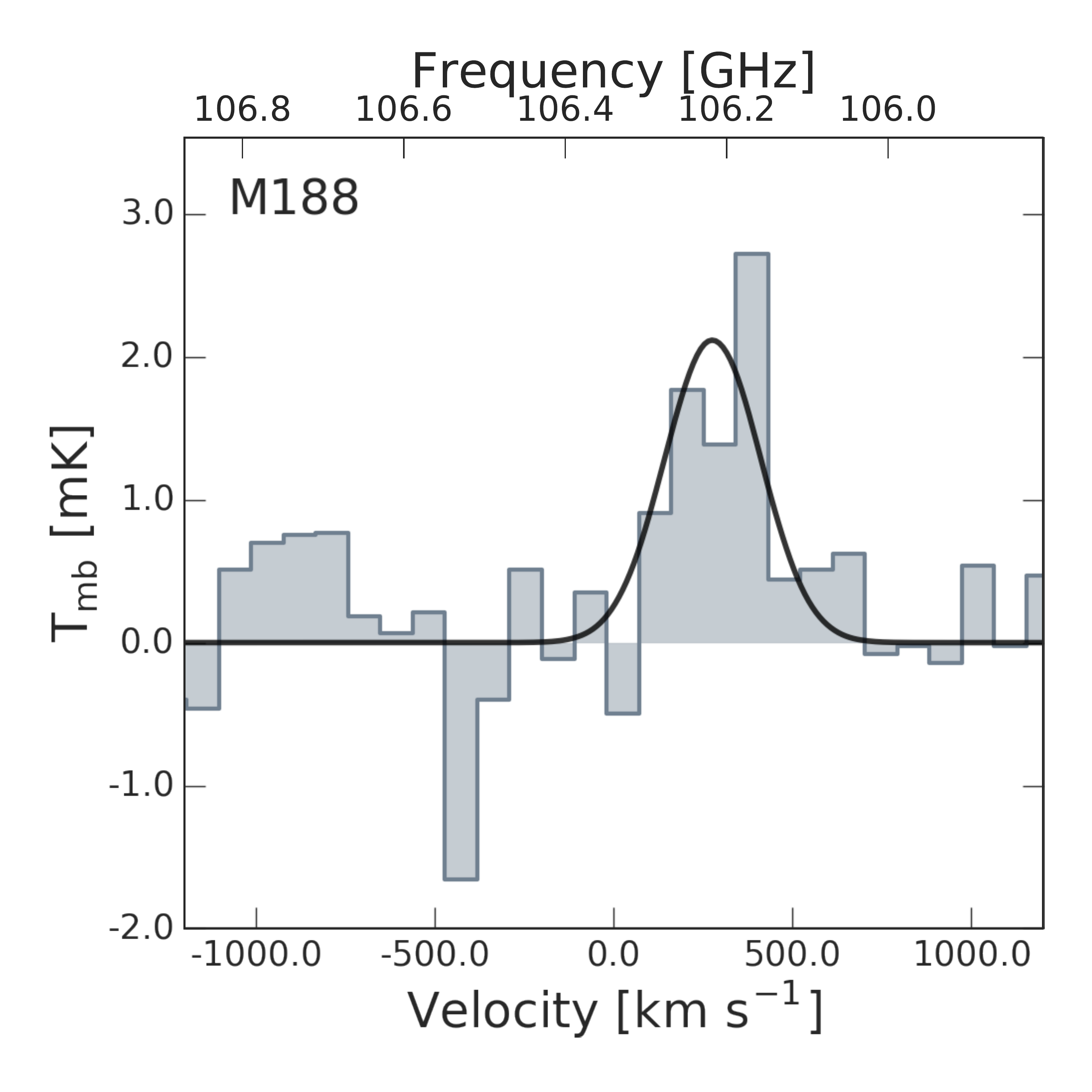}
  \end{subfigure}
  \begin{subfigure}{0.3\textwidth}
	  \includegraphics[width=\linewidth]{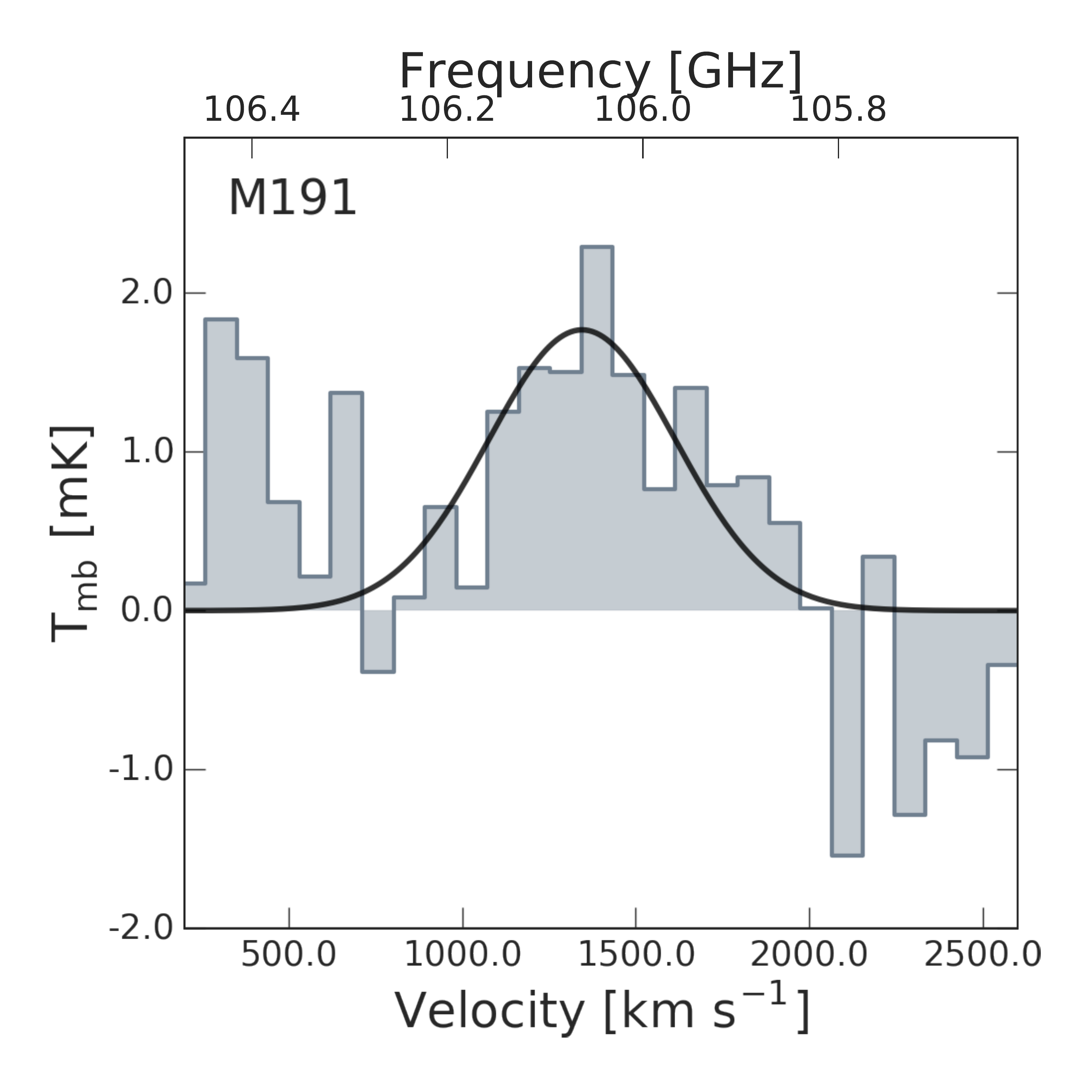}
  \end{subfigure}
  \caption{(Continued)} 
\end{figure*}

\begin{table*}
\centering
\caption{CO(1--0) line observations of the 5MUSES targets from our programs.}
\label{tab:muses_co}
\begin{tabular}{ccclcccc} 
\hline
\hline
${\rm ID}^{\rm a}$ & R.A. & Decl. & $z_{\rm co}$ & FWHM$^{\rm e}$ & ${\rm S}_{\rm CO} \Delta v$ & log(\lco$^{\rm d}$) & rms$^{\rm f}$ \\
& [hh:mm:ss] & [dd:mm:ss] & & [\si{\km\per\s}] & [\si{\Jy\km\per\s}]  & & [mK] \\
\hline
20  & 02:19:09.6 & --05:25:12.9   & 0.098$^{\rm b}$ & --  & \textless0.16  & \textless7.87 & 0.94     \\
22  & 02:19:16.1 & --05:57:27.0     & 0.103  & $311\pm84$  & $1.79\pm0.47$  & $8.95\pm0.11$  & 0.30   \\
28  & 02:19:53.0 & --05:18:24.2& 0.073 & $261\pm73$  & $5.94\pm1.40$  & $9.16\pm0.10$ & 0.61    \\
36  & 02:21:47.9 & --04:46:13.5& 0.025$^{\rm b}$ & --  & \textless0.65  & \textless7.27 & 0.57   \\
64  & 02:25:48.2 & --05:00:51.5& 0.150$^{\rm c}$ & --  & \textless0.31  & \textless8.52  & 0.29  \\
66  & 02:26:00.0   & --05:01:45.3   & 0.205 & $274\pm59$  & $2.37\pm0.45$  & $9.69\pm0.08$  & 0.16  \\
75  & 02:27:41.6 & --04:56:50.6   & 0.055 & $173\pm61$   & $3.27\pm0.85$   & $8.66\pm0.11$ & 0.07   \\
86  & 10:36:46.4 & +58:43:30.6 & 0.140 & $307\pm80$  & $2.58\pm0.71$  & $9.38\pm0.12$  & 0.28  \\
106 & 10:44:38.2 & +56:22:10.8& 0.024 & $375\pm68$  & $15.22\pm2.38$   & $8.64\pm0.07$  & 1.31  \\
107 & 10:44:54.1 & +57:44:25.8   & 0.118 & $345\pm74$  & $3.59\pm0.79$   & $9.37\pm0.10$  & 0.21  \\
118 & 10:49:07.2 & +56:57:15.4& 0.071 & --  & \textless3.55 & \textless8.93 & 0.80    \\
123 & 10:50:06.0   & +56:15:00.0  & 0.118 & $215\pm56$  & $2.65\pm0.63$  & $9.25\pm0.10$ & 0.21   \\
146 & 10:59:03.5 & +57:21:55.1& 0.117$^{\rm b}$ & --  & \textless0.87  & \textless8.76 & 0.51  \\
152 & 11:01:33.8 & +57:52:06.6   & 0.275 & $285\pm89$  & $3.71\pm1.15$  & $10.15\pm0.13$  & 0.24 \\
185 & 16:08:58.4 & +55:30:10.3   & 0.065 & $221\pm38$  & $3.40\pm0.66$   & $8.84\pm0.08$ & 0.37   \\
187 & 16:09:07.6 & +55:24:28.4   & 0.065           & $396\pm300$  & $5.28\pm1.07$   & $9.01\pm0.09$  & 0.39  \\
188 & 16:09:08.3 & +55:22:41.5& 0.085           & $313\pm73$  & $5.28\pm1.07$  & $9.01\pm0.09$ &  0.54   \\
191 & 16:09:31.6 & +54:18:27.4   & 0.086           & $645\pm165$ & $7.58\pm1.67$  & $9.38\pm0.10$ & 0.36     \\
192 & 16:09:37.5 & +54:12:59.3& 0.086           & $294\pm136$ & $3.28\pm0.64$  & $9.06\pm0.08$  & 0.13  \\
196 & 16:12:23.4 & +54:03:39.2   & 0.138           & $210\pm65$   & $1.93\pm0.53$   & $9.24\pm0.12$ & 0.18    \\
197 & 16:12:33.4 & +54:56:30.5& 0.084           & $279\pm38$  & $7.76\pm1.13$  & $9.40\pm0.06$ & 0.88   \\
198 & 16:12:41.1 & +54:39:56.8& 0.035$^{\rm b}$ & --  & \textless1.23  & \textless7.84  & 0.61  \\
200 & 16:12:50.9 & +53:23:05.0& 0.047           & -- & \textless5.50  & \textless8.77  & 0.85  \\
202 & 16:12:54.2 & +54:55:25.4& 0.065           & --  & \textless5.48  & \textless9.03   & 0.99 \\
294 & 17:12:32.4 & +59:21:26.2& 0.210            & $307\pm55$   & $2.00\pm0.26$  & $9.64\pm0.06$ & 0.03   \\
297 & 17:13:16.6 & +58:32:34.9& 0.079           & $437\pm101$ & $3.62\pm0.97$ & $9.02\pm0.12$ & 0.33   \\
302 & 17:14:46.4 & +59:33:59.8   & 0.131           & --  & \textless3.49  & \textless9.44 &  0.75  \\
310 & 17:17:11.1 & +60:27:10.0& 0.110            & $350\pm99$   & $3.15\pm0.79$  & $9.26\pm0.11$ & 0.17   \\
315 & 17:19:33.3 & +59:27:42.7   & 0.139           & $423\pm131$ & $3.85\pm1.05$  & $9.55\pm0.12$  & 0.42 \\
316 & 17:19:44.9 & +59:57:07.1& 0.069           & --  & \textless2.15  & \textless9.03 &  1.36  \\
317 & 17:20:43.3 & +58:40:26.9   & 0.125           & $329\pm152$  & $2.81\pm0.73$   & $9.32\pm0.11$ & 0.11   \\
319 & 17:21:59.3 & +59:50:34.2& 0.028           & $168\pm46$  & $7.67\pm1.96$  & $8.44\pm0.11$ & 0.51   \\
328 & 17:25:46.8 & +59:36:55.3   & 0.035           & $393\pm137$    & $13.89\pm2.53$ & $8.89\pm0.08$  &  0.24 \\
329 & 17:25:51.3 & +60:11:38.9   & 0.029           & $171\pm35$  & $9.10\pm1.67$  & $8.54\pm0.08$  &  0.04 \\ \hline \hline
\end{tabular}
\begin{flushleft}
$^{\rm a}$5MUSES ID name.\\
$^{\rm b,c}$For sources with CO(1--0) 3$\sigma$ upper limits, the redshift is derived from the IRS spectra (b) or obtained from the NASA/IPAC Extragalactic Database (c) as listed in \citet{Wu2010}.\\
$^{\rm d}$\lco\ luminosities are in units of [\ulco].\\
$^{\rm e}$The CO line width is estimated by measuring the full width at half-maximum (FWHM) of the Gaussian profile.\\
$^{\rm f}$ For galaxies with $>3\sigma$ CO detection and upper limits, we list the RMS of the CO line and the baseline, respectively.
\end{flushleft}
\end{table*}

\begin{table*}
\centering
\caption{Data sample overview.}
\label{tab:overview}
\begin{tabular}{llcclll}
\hline
\hline
Sample       & $N$ & $z$      & log(\lir/\lsol)  & PAH feature & CO line  & References [PAH, CO] \\ \hline
SINGS        & 36 & 0.001--0.007 & 7.31--10.60  & 6.2 \si{\micro\metre}  & (3-2)   & S07, W12 \\
Local ULIRGs & 9, (9) & 0.018--0.191  & 11.99--12.42    & 6.2 \si{\micro\metre}, 7.7 \si{\micro\metre}  & (1-0)& P13  \\
5MUSES       & 22, (24) & 0.025--0.277  & 9.23--11.81  & 6.2 \si{\micro\metre}, 7.7 \si{\micro\metre} & (1-0) & W10, this paper \\
5MUSES       & 14, (15) & 0.053--0.360  & 10.35--12.10    & 6.2 \si{\micro\metre}, 7.7 \si{\micro\metre} & (1-0)  & W10, K14 \\
high-$z$ SFGs  & 4, (3) & 1.016--1.523  & 11.86--12.66& 6.2 \si{\micro\metre}, 7.7 \si{\micro\metre}  & (1-0), (2-1) & P13   \\
SMGs & 6, (8) & 1.562--4.055  & 12.30--13.08 & 6.2 \si{\micro\metre}, 7.7 \si{\micro\metre}  & (1-0), (2-1), (3-2), (4-3) & P13 \\
High-$z$ SBs  & (6) & 1.562--2.470  & 12.55--12.93 & 6.2 \si{\micro\metre}, 7.7 \si{\micro\metre}  & (2-1), (3-2) & Sa07, Y07 \\ \hline \hline
\end{tabular}
\begin{flushleft}
\textbf{Notes} -- The table includes all galaxies with at $3\geqslant\sigma$ CO and PAH detections. \\
Col. (1): Galaxy sample; col. (2): Number of galaxies with PAH 6.2 \si{\micro\metre} (and 7.7 \si{\micro\metre}) detections; col. (3): Redshift range; col. (4): IR luminosity range; col. (5): Detected PAH feature; col. (6): Observed CO transition line; col. (7): References: S07: \citet{Smith2007}, W12: \citet{Wilson2012}, P13: \citet{Pope2013} and references therein, W10: \citet{Wu2010}, K14: \citet{Kirkpatrick2014}, Sa07: \citet{Sajina2007}, Y07: \citet{Yan2007}. \\
\end{flushleft}
\end{table*}

\subsection{Literature data}
To expand our data sample, we include published observations of galaxies at all redshifts with both CO, IR, and PAH detection from the literature. General properties of the data compilation are listed in Table \ref{tab:overview}.

\subsubsection{5MUSES galaxies}
In a recent study, \citet{Kirkpatrick2014} carried out CO(1--0) line observations of 24 intermediate redshift galaxies ($z=0.04-0.36$), also selected from the 5MUSES sample, with the Redshift Search Receiver (RSR) on the Large Millimetre Telescope (LMT). Their sample covers a broader range of \lir\ ($10^{10.4}-10^{12.1}$ \lsol) and \ew\ ($0.07-0.70$ \si{\micro\metre}) as opposed to our targets. They detected CO(1--0) emission in 17 of the 24 sources which we combine with our sample for the analysis (14 of these have $3\sigma$ PAH 6.2 \si{\micro\metre} detection). For consistency, we derive CO line luminosities using the velocity integrated line flux reported in \citet{Kirkpatrick2014}. The final sample of intermediate redshift galaxies with both CO and PAH 6.2 \si{\micro\metre} detections in our study consists of 36 targets, all drawn from the 5MUSES compilation, of which 24 are from our new IRAM survey and 14 from \citet{Kirkpatrick2014}.

\subsubsection{SINGS}\label{sec:sings}
The \s\ Infrared Nearby Galaxy Survey \citep[SINGS:][]{Kennicutt2003} is an imaging and low-resolution ($R \sim 50-100$) spectroscopic survey of 75 local galaxies with $5-38$ \si{\micro\metre} spectral mapping with \s\ IRS. The low-resolution $5-15$ \si{\micro\metre} spectral map ($55'' \times 34 ''$) is centred on the nucleus of each galaxy. From the SINGS sample, \citet{Smith2007} selected 59 galaxies with spectral coverage between $5-38$ \si{\micro\metre} from both SL and LL in order to detect PAH emission within the central regions of each galaxy. PAH features were derived using PAHFIT and fitted with Drude profiles.  A subsample of 57 SINGS galaxies have both PAH emission and FIR coverage based on \h\ PACS and SPIRE observations as presented in \citet{Dale2012}. 

From the SINGS survey, \citet{Wilson2012} selected 47 galaxies (NGLS: Nearby Galaxies Legacy Survey) to carry out CO(3--2) line observations with the James Clerk Maxwell Telescope (JCMT). To correct the CO(3--2) emission to CO(1--0) we adopt a CO(3--2)/CO(1--0) line ratio of $r_{32/10}=0.18\pm0.02$ based on a comparison study by \citet{Wilson2012}. They estimate an average CO(3--2) and CO(1--0) line ratio using 11 nearby galaxies from the NGLS sample that overlap with CO(2--1) observations carried out by \citet{Kuno2007}. We apply an aperture correction ($f_{\rm TIR}$) listed in \citet{Smith2007} to the IR and CO(1--0) luminosities and increase the uncertainties of the aperture-corrected luminosities by a factor of two, in order to compare these with the PAH emission arising from the central part of the galaxy. 

Given the different physical scales probed by the available PAH, CO, and dust emission observations of the SINGS galaxies, we choose to exclude them from our statistical analysis to avoid biases in our regression models due to possible systematics and uncertainties introduced by the aperture corrections. However, since the SINGS sample consists of representative, normal PAH emitting SFGs in the local universe, for the sake of completeness we choose to overplot them in the various luminosity scaling relations presented in this study.

\subsubsection{Local ULIRGs and high$-z$ galaxies}
\citet{Pope2013} carried out CO(2--1) observations using IRAM PdBI and \s\ MIR spectroscopy of six 70 \si{\micro\metre} selected galaxies from the \s\ Far-Infrared Deep Extragalactic Legacy survey \citep[][]{DickinsonFIDEL2007} of GOODS-N with optical spectroscopic redshift at $z=1-1.5$. The sample has \s\ IRS observations and photometric coverage from \s\ MIPS (24 and 70 \si{\micro\metre}) and \h\ PACS (100 and 160 \si{\micro\metre}) and SPIRE (250, 350 and 500 \si{\micro\metre}) observations from the GOODS-\h\ survey \citep{Elbaz2011}. As in \citet{Pope2013}, we complement our sample with galaxies from the literature containing detected CO and PAH emission at all redshifts. These include 12 high-$z$ galaxies (SMGs, BzKs, and 70 \si{\micro\metre} selected galaxies) from various studies at $1.1<z<4.1$ \citep{Pope2008b, Frayer2008, Aravena2010, Carilli2010, Casey2011, Ivison2011, Magnelli2012, Bothwell2013, Riechers2013}. Stellar masses from \citet{Pope2013} are available for the 70 \si{\micro\metre} selected galaxies. Based on the offset from the MS, we classify the 70 \si{\micro\metre} as high-$z$ SFGs galaxies whereas the remaining galaxies at $z>1$ are labelled as SMGs.
Similarly, we include the 24 \si{\micro\metre} selected sample ($S_{24} > 0.9$\,mJy) of 9 $z\sim 1-2$ ULIRGs in the \s\ XFLS field with CO(2--1) or CO(3--2) observations from \citet{Yan2010} that also has existing \s\ MIR spectra published in \citet{Yan2007} and \citet{Sajina2007}.

For galaxies with only high-$J$ CO line observations, we convert the CO luminosities to \lco\ by adopting the conversion factors listed in \citet[][$r_{21/10}=0.84\pm0.13$, $r_{32/10}=0.52\pm0.09$, $r_{43/10}=0.41\pm0.07$]{Bothwell2013}. At lower redshift, we also include 13 local ULIRGs with \s\ IRS MIR spectra from \citet{Armus2007} and \citet{Desai2007} with existing CO observations and IR luminosities \citep{Sanders1991, Solomon1997, Kim1998, Farrah2003, GaoSolomon2004, Chung2009}. The CO, IR, and PAH luminosities of the literature compilation are available in the online version (See Table \ref{tab:literature}).

\subsection{Derivation of MIR and FIR dust properties}\label{sec:sed}
We combine the existing multi-wavelength photometry from \s\ (MIPS: 24, 160 \si{\micro\metre}) and \h\ (SPIRE: 250, 350, 500 \si{\micro\metre}) in order to estimate IR luminosities and dust masses (\md) by modeling the FIR part of the spectral energy distribution (SED) for each galaxy in the 5MUSES sample. We use silicate-graphite-PAH models from \citet{Draine2007} (DL07) including diffuse ISM and photodissociation region (PDR) components. The best-fit parameters and results are listed in the online version (see Table \ref{tab:muses_all}).
\lir\ is derived by integrating the SED model between rest-frame $8-1000$ \si{\micro\meter}. The SFR for each galaxy is estimated using the \lir--SFR conversion in  \citet{Kennicutt1998} assuming a \citet{Salpeter1955} IMF: ${\rm SFR}$ [\sfr]$=1.72\times 10^{-10}$ \lir. This technique of FIR SED modeling is applied to the 5MUSES sample including 165 galaxies. In addition, we estimate monochromatic dust luminosities using the \s\ MIPS (24 and 160 \si{\micro\metre}) and \h\ SPIRE photometric bands (250, 350, and 500 \si{\micro\metre}). We derive PAH luminosities from the \s\ IRS data using PAHFIT (\citet{Smith2007}). For galaxies in the literature where the PAH luminosities have been estimated using the spline method, we derive the PAH luminosities using PAHFIT to ensure that the PAH 6.2 and 7.7 \si{\micro\metre} emissions have been estimated in a consistent way for both our targets and the literature compilation.

\noindent
To summarize, the full sample with both detected CO, PAH 6.2 \si{\micro\metre}, and IR emission contains 36 5MUSES galaxies (including 5 AGNs and composite sources), 36 SINGS galaxies, 9 local ULIRGs, 4 high-$z$ SFGs, and 6 SMGs (See Table \ref{tab:overview}). 

\begin{figure*}
\begin{subfigure}{0.46\textwidth}
\includegraphics[width=\linewidth, trim={0 0 0 0}, clip]{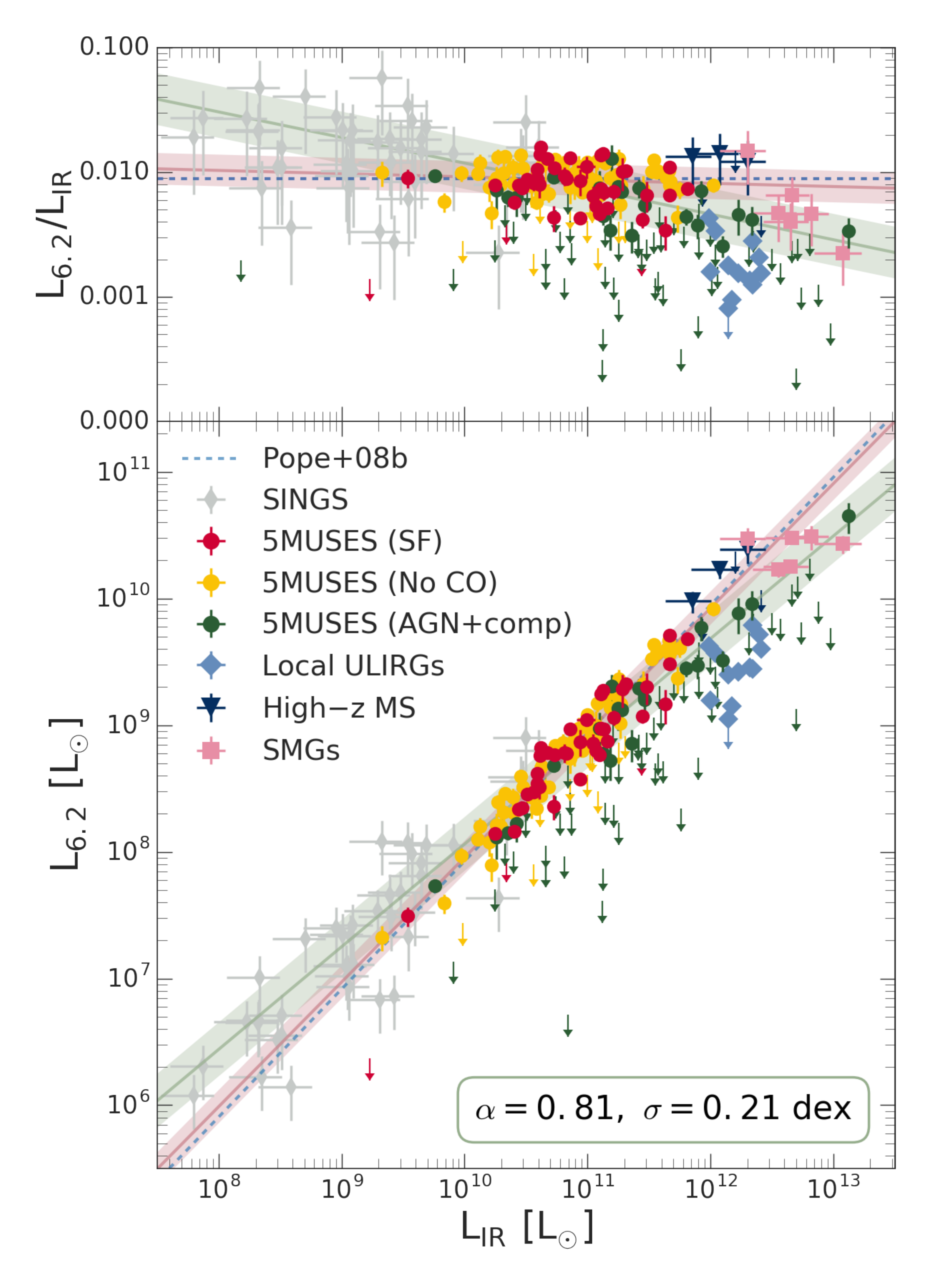}
\end{subfigure} \hfill
\begin{subfigure}{0.46\textwidth}
\includegraphics[width=\linewidth, trim={0 0 0 0}, clip]{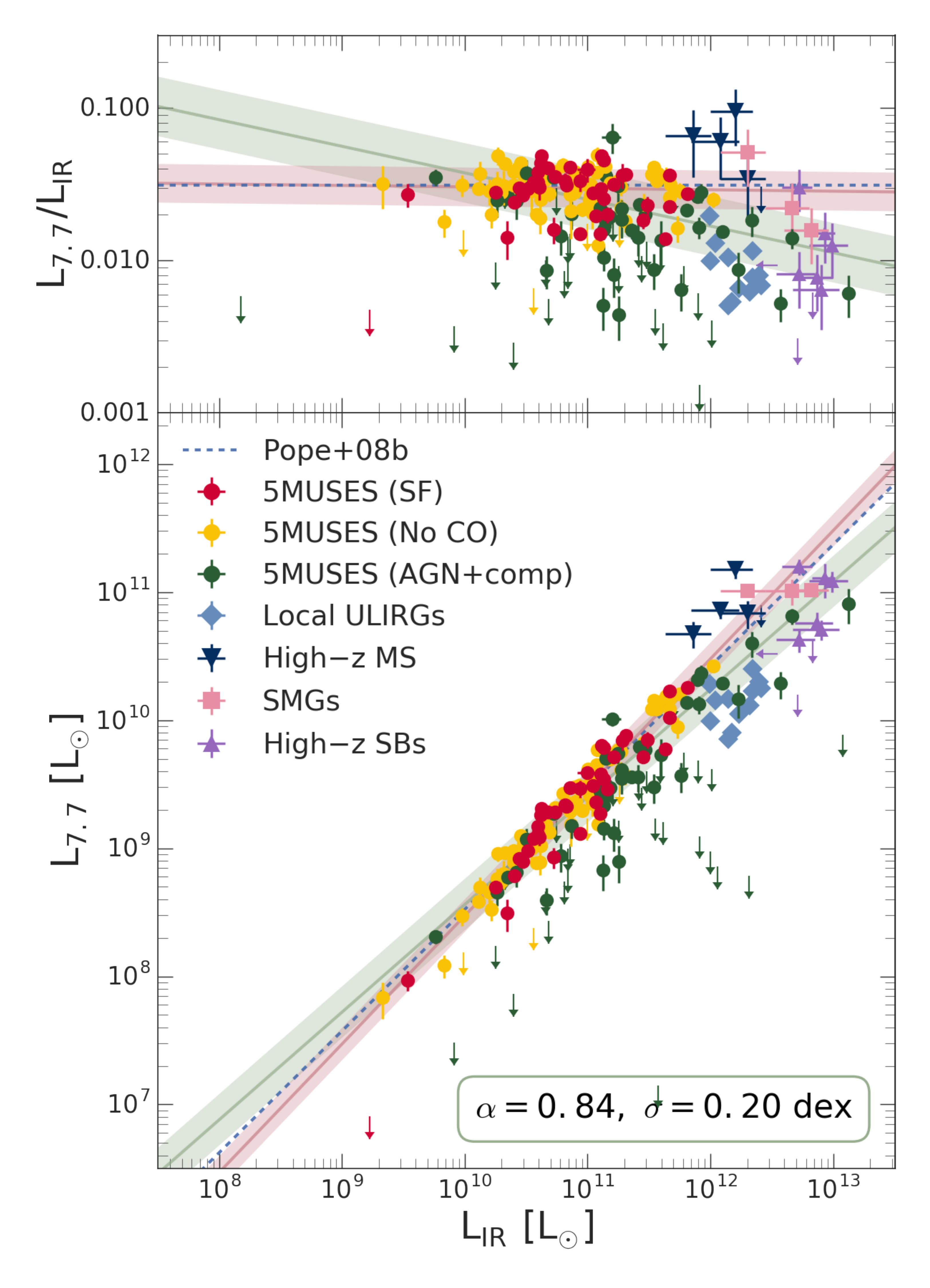}
\end{subfigure}
	\caption{Correlation between the total infrared luminosity (\lir) vs. PAH $6.2$ \si{\micro\metre} luminosity (\lpah) (\textit{left}) and PAH 7.7 \si{\micro\metre} luminosity (\lpahsev) (\textit{right}). For the 5MUSES sample, we include CO-detected SFGs (red), AGNs and composite sources (green), and the remaining sample of SFGs (yellow). \lpah\ and \lpahsev\ upper limits are shown as arrows. We  also includes SINGS galaxies (grey, only for \lpah), local ULIRGs (light blue), high-$z$ SFGs (dark blue), SMGs (pink), and high$-z$ SBs (purple, only for \lpahsev).
	The green and pink lines depict the \lpahall--\lir\ linear regression models of the SFGs (5MUSES and high-$z$ SFGs) and the full sample, respectively, excluding the SINGS sources as described in Section \ref{sec:sings}. The same method is applied to the \lpahall/\lir\ vs. \lir\ relations presented in the upper panels. The shaded regions present the intrinsic scatter of the best-fits. The blue dashed lines show the best-fit relations from \protect\citet{Pope2008b} of local SBs and SMGs. For the upper panels, the dashed lines are the median value of the \lpahall/\lir\ assuming \lpahall-\lir\ 
    slopes of unity.}    
    \label{fig:lir_lpah}
\end{figure*}

\section{Results}
Previous works have studied scaling relations between the \lir, \lpahall, and \lco\ of various galaxy populations across a wide range of redshifts \citep[e.g..][]{Calzetti2005, Smith2007, Bendo2008, Pope2013, Rujopakarn2013, Kirkpatrick2014}. In this section, we will revisit these relations for our sample, attempting to identify outliers and investigate them not only as a function of look-back time but also as a function of physical conditions (AGN, SBs, normal galaxies, etc).

\begin{figure}
	\centering
	\includegraphics[width=0.47\textwidth, trim={0 0 0 0}, clip]{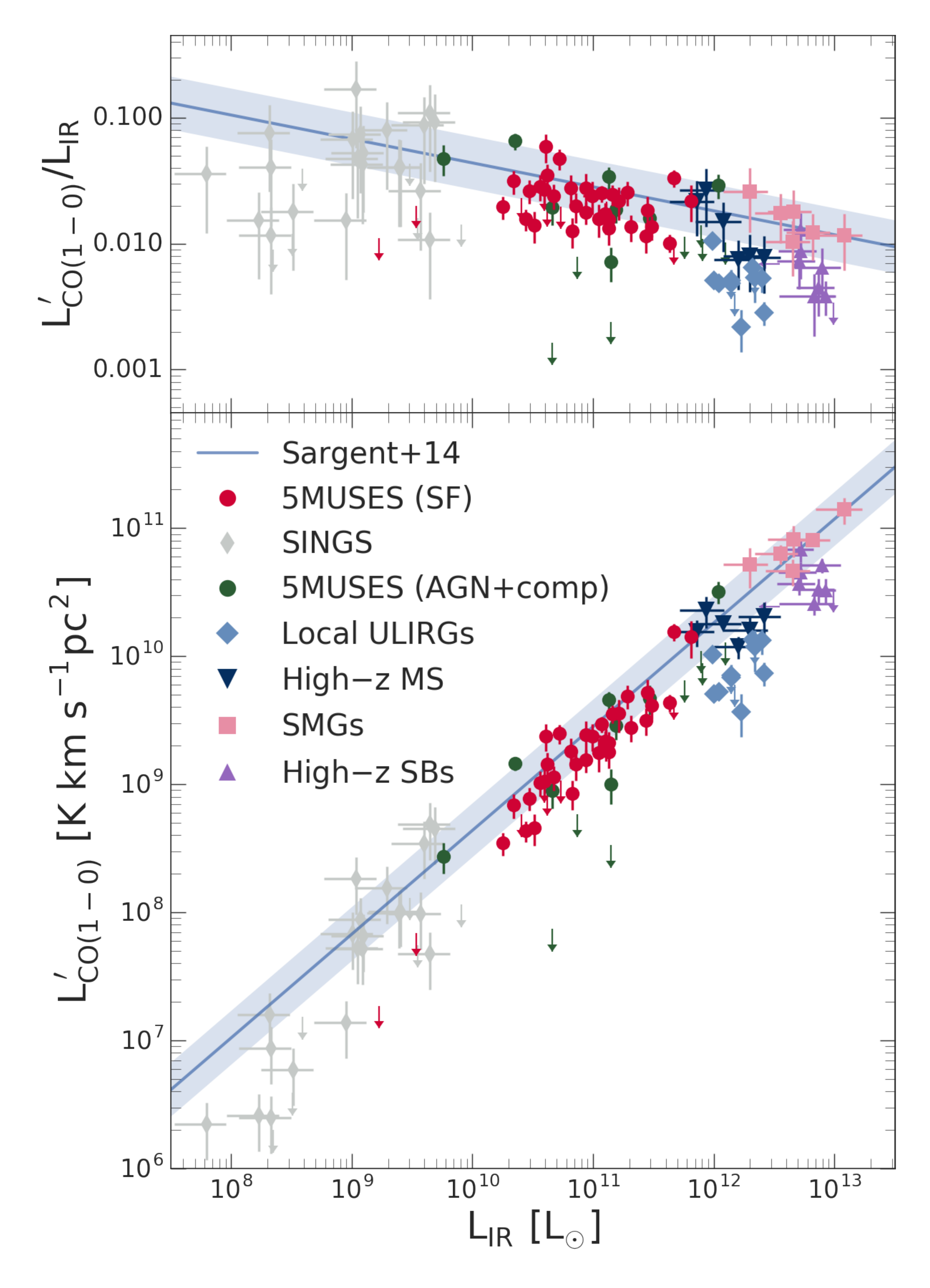}
    \caption{Correlation between CO(1--0) line luminosity (\lco) and \lir. Colour coding and symbols follow Figure \ref{fig:lir_lpah}. Higher-\textit{J} CO transitions from the literature are corrected to CO(1--0) using \protect\citet{Wilson2012} for the SINGS galaxies and \protect\citet{Bothwell2013} are adopted for the remaining sample. The blue lines shows the \lco--\lir\ relation from \protect\citet{Sargent2014}. The top panel shows the \lco/\lir\ ratio in units of [(\ulco)/\lsol] as a function of \lir\ along with the \protect\citet{Sargent2014} relation. The blue shaded regions depict the observed dispersion of 0.21 dex.}
    \label{fig:lco_lir}
\end{figure}

\subsection{The relation between IR and PAH luminosity} \label{sec:pah_ir}
As discussed in the Introduction, \lpahall/\lir\ variations could be driven by the presence of an AGN,  the geometry of the star formation but also by the metallicity and the hardness of the radiation field. To investigate the \lpahall/\lir\ variations in different galaxy populations across a wide range of redshifts, we consider the \lpah, \lpahsev, and \lir\ measurements for the 5MUSES sample as well as for other galaxies in the literature for which such measurements are available, including local ULIRGs \citep{Armus2007, Desai2007}, high$-z$ ($1<z<4$) SMGs, BzKs and 70 \si{\micro\metre} selected galaxies \citep[][and references therein]{Pope2008b, Pope2013}, and 24 \si{\micro\metre} selected SBs at $z \sim2$ \citep{Yan2010}. Due to possible biases introduced from the aperture correction applied to the SINGS galaxies \citep{Kennicutt2003}, we omit them from the best-fitted regression models.

In Figure \ref{fig:lir_lpah}, we plot the PAH 6.2 \si{\micro\metre} and 7.7 \si{\micro\metre}  luminosities as a function \lir\ for our full sample of galaxies and model the data in the logarithmic space using the Bayesian linear regression analysis as described in \citet{Kelly2007}. This method accounts for measurement errors of both the dependent and independent variables and it returns posterior distributions of the best-fit parameters, including the intrinsic scatter. All the best-fit parameters of the regression model: ${\rm log}~y = \alpha \times {\rm log}~x + \beta$ are listed in Table \ref{tab:scaling}. Focusing only on local/intermediate redshift star-formation dominated sources and high$-z$ SFGs that are part of the MS of star formation, we find a tight, linear correlation, with a slope of $\alpha=0.98\pm0.03$ and intrinsic scatter of $\sigma=0.13$ dex for the \lpah--\lir\ relation, while $\alpha=1.00\pm0.03$ and $\sigma=0.13$ dex for \lpahsev--\lir, in agreement with the best-fit relations reported in \citet{Pope2008b} and similar studies \citep[e.g..][]{Sajina2008, Rujopakarn2013, Shipley2016}. The dispersion of the \lpahall/\lir\ ratios as a function of \lir\ and galaxy type is shown in the top panels of Figure \ref{fig:lir_lpah}. From these relations, local ULIRGs (and high-$z$ SBs with available \lpahsev\ estimates) exhibit systematically lower PAH/IR luminosity ratios. We quantify these galaxies as outliers lying $3.2\sigma$ from the best-fit relations ($2.5\sigma$ for \lpahsev--\lir). Fitting the full sample (excluding upper limits), yields a shallower slope but also an increased intrinsic scatter of $\sigma=0.21$ dex and $\sigma=0.20$ dex for the \lpah--\lir\ and \lpahsev--\lir\ relations, respectively. We note though, that a considerable fraction of local ULIRGs and high-$z$ SB are still outliers, even when attempting to fit the whole sample, in agreement with \citet{Shipley2016}. It is thus evident that a universal \lir--\lpah\ relation, accommodating the various physical conditions of different galaxy populations, cannot be established. 

\begin{figure*}
\begin{subfigure}{0.46\textwidth}
	\includegraphics[width=\linewidth, trim={0 0 0 0}, clip]{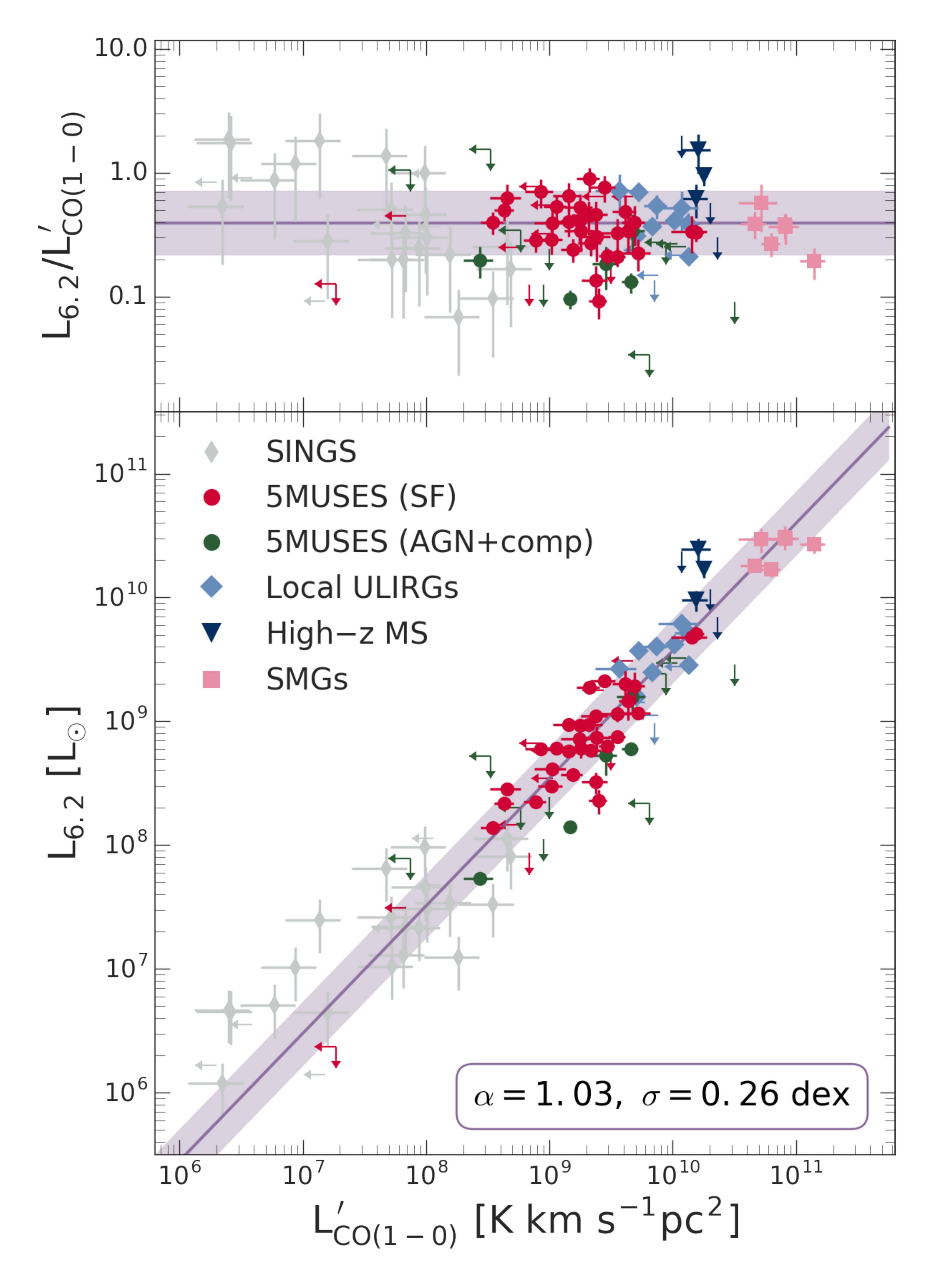} 
\end{subfigure} \hfill
\begin{subfigure}{0.46\textwidth}
	\includegraphics[width=\linewidth, trim={0 0 0 0}, clip]{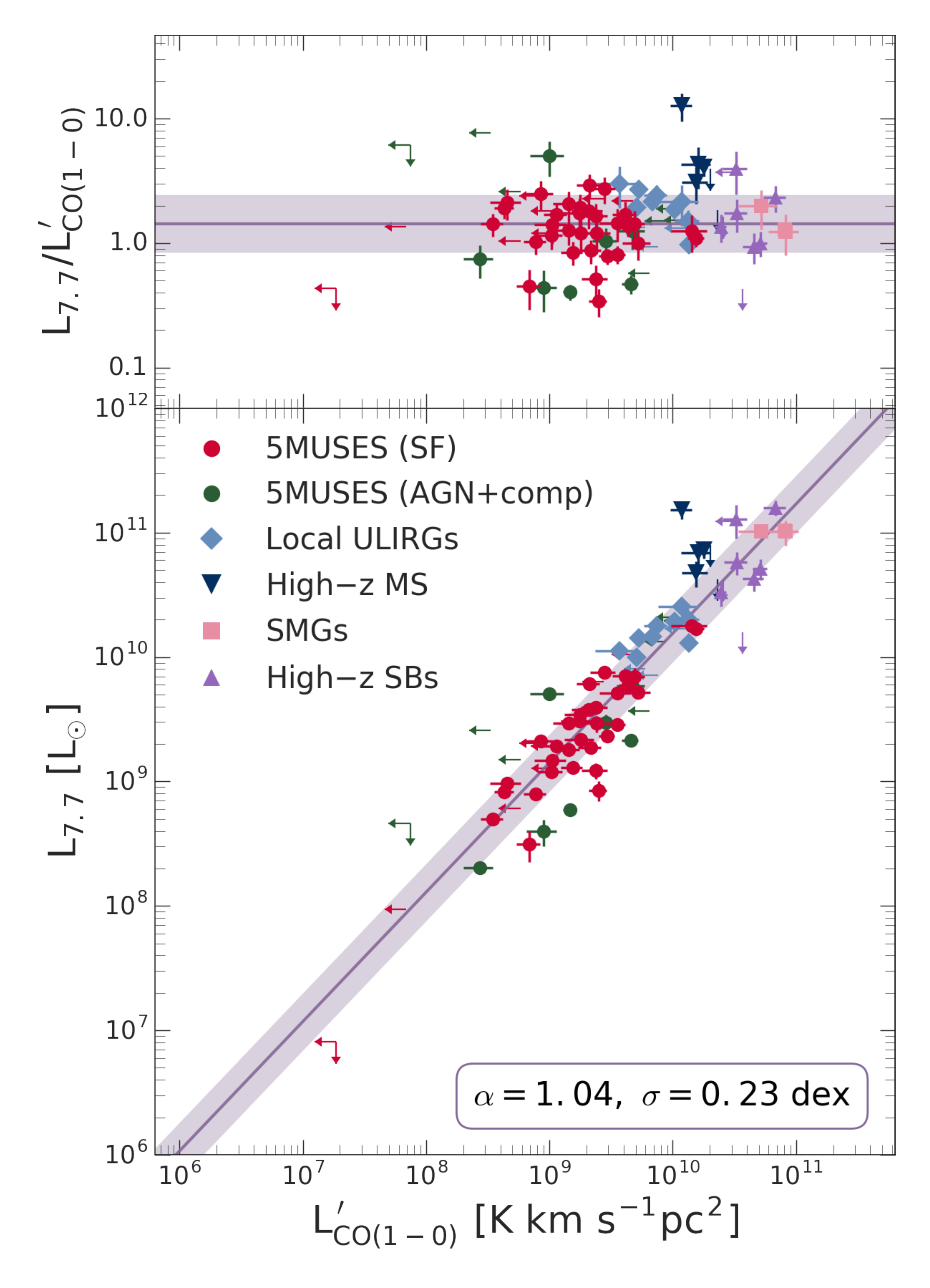} 
\end{subfigure}
	\caption{Correlation between \lpah\ vs. \lco (\textit{left}) and \lpahsev\ vs. \lco\ (\textit{right}). Colour coding and symbols follow Figure \ref{fig:lir_lpah}. 
    The top panels show the PAH and CO luminosity ratios in units of [\lsol/(\ulco)] and as a function of \lco.  SINGS galaxies (grey) are excluded from the fitting procedure due to possible  systematics introduced from the applied aperture correction (See Section \ref{sec:sings}). The purple lines and shaded regions depict the best-fit linear regression and its intrinsic scatter for all the galaxies. For the upper panels, we show the observed dispersion assuming a \lpahall--\lco\ slope of unity. The fit parameters of the CO--PAH   luminosity relations are listed in Table \ref{tab:scaling}.}
	\label{fig:fig1}
\end{figure*}

\subsection{The CO--IR luminosity relation} \label{sec:co_ir}
The use of \lir\ as a SFR tracer and the fact that CO emission is directly associated with the molecular gas reservoir of a galaxy, has motivated several studies to investigate the \lco--\lir\ relation as a proxy of the star formation law that links the SFR to the molecular gas of galaxies (KS law). The existence of a universal \lco--\lir\ (and thus of a universal \mgas--SFR) relation has been challenged by recent observations of different galaxy populations at various redshifts \citep[e.g..][]{Bouche2007, Daddi2010b, Genzel2010, Krumholz2012, Silverman2015}. Although the debate is still open, there are claims that MS galaxies at all redshifts tend to follow a unique \lco--\lir\ relation from which local ULIRGs and high$-z$ SBs are outliers exhibiting lower \lco/\lir\ ratios, indicative of higher SFEs \citep[e.g..][]{Daddi2010b, Genzel2010}. Addressing the question of this possible bimodality is beyond the scope of our work. Instead, we wish to investigate how the galaxies at different redshifts and with different physical conditions populate the \lco--\lir\ parameter space and explore how the global \lco--\lir\ relation behaves with respect to the observed trends between \lpahall--\lir\ and \lpahall--\lco.

In Figure \ref{fig:lco_lir}, we present the \lir--\lco\ relation for the 5MUSES sample along with the literature compilation included in Figure \ref{fig:lir_lpah} (See Table \ref{tab:overview}). We also consider the \lir--\lco\ relation of \citet{Sargent2014} calibrated on MS galaxies with \Mstar$\geqslant 10^{10}$ \msol\ at $0 < z < 3.2$ with a slope of 0.81 and a dispersion of 0.21 dex. The vast majority of SFGs, including our 5MUSES sample and the high-$z$ SFGs appear to follow the Sargent relation. On the other hand, local ULIRGs and high-$z$ SBs are outliers, a situation that resembles the \lpah--\lir\ and \lpahsev--\lir\ relations (Figure \ref{fig:lir_lpah}). In other words, sources exhibiting lower \lco/\lir\ ratios with respect to the general population of normal galaxies, tend to also exhibit lower \lpahall/\lir\ ratios. In the next subsection, we bring these two together by exploring the relations between PAH and CO luminosities.

\subsection{The relations between CO and PAH emission}\label{sec:co_pah}
In Figure \ref{fig:fig1}, we plot the \lpah\ vs. \lco\ (\textit{left}) and \lpahsev\ vs. \lco\ (\textit{right}) luminosity relations for the 5MUSES sample as well as for the whole data set confirming the observed correlation between the PAH 6.2 \si{\micro\metre} and CO emission \citep{Pope2013}. The various populations appear to follow a unique relation, with a slope of unity within the uncertainties (\lpah--\lco: $\alpha = 1.03 \pm 0.06$ and \lpahsev--\lco: $\alpha = 1.04 \pm 0.08$), in agreement with the slightly sub-linear slope ($\alpha=0.9\pm0.01$) reported in \citet{Pope2013}. 
The \lpah--\lco\ and \lpahsev--\lco\ relations have an intrinsic scatter of $\sigma=0.26$ dex and $\sigma=0.23$ dex respectively,  without any specific star-forming galaxy population standing out as prominent outliers.
The shaded regions in the upper panels of Figure \ref{fig:fig1} depict the intrinsic scatter of \lpah--\lco\ and \lpahsev--\lco\ correlations assuming a slope of unity. In order to ensure that the linear slopes of the \lpahall--\lco\ correlations are not affected by corrections applied to galaxies with higher-\textit{J} CO observations, we fit only those galaxies with CO(1--0) emission yielding a slope of $\alpha=0.99\pm0.07$ consistent with the best-fit of the full sample. The global PAH-CO luminosity relation for \lpah\ is parametrized as:
\begin{eqnarray} \label{eq:lco_l62}
L_{\rm 6.2}\,\,\, [\rm L_{\sun}] = (0.39 \pm 0.18) \times L_{\rm CO}^{\prime}\,\, 
\rm [K\,\, km\,\, s^{-1}\,\, pc^{2}]
\end{eqnarray}

\noindent The universal \lpah--\lco\ and \lpahsev--\lco\ relations as indicated by our data suggest a link between the PAH and CO emission that appears unaffected by the physical conditions of the galaxies. Since the CO emission is a tracer of the gas mass and thus of the cold dust emission, a natural consequence of the \lpahall--\lco\ is a link between the PAH and the cold dust emission of a galaxy. This is explored in detail in the following Section.

\subsection{The relation between PAH and dust emission}\label{sec:pah_dust}
In the previous section, we showed that the emission from PAHs on global scales correlate with CO(1--0) luminosity over a wide range of redshifts and various galaxy types. This result suggests a link between the PAH emission and the \mgas\ of a galaxy. Since \mgas\ is a derived physical parameter rather than a direct observable, before exploring a possible \lpahall--\mgas\ relation, it is informative to investigate the scaling relations and the scatter between \lpahall\ and the warm and the cold dust emission through MIR and FIR photometric bands. The motivation behind this exercise is that \mgas\ is known to be directly associated with the cold dust emission of galaxies \citep{Leroy2011, Eales2012, Magdis2012, Magdis2013, Scoville2017}, whereas the warm dust emission is linked to star formation. For the sake of brevity and clarity, we only present the results for the PAH 6.2 \si{\micro\metre} feature which is least affected by silicate absorption and extinction \citep{Peeters2004}. However, the same applies to the PAH 7.7 \si{\micro\metre} feature due to the linear correlation between these two in logarithmic scales (\lpah--\lpahsev: $\alpha=1.02\pm 0.02$, see Table \ref{tab:scaling}).

\begin{figure*}
	\centering
	\includegraphics[width=0.7\textwidth, trim={0 0 0 0}, clip]{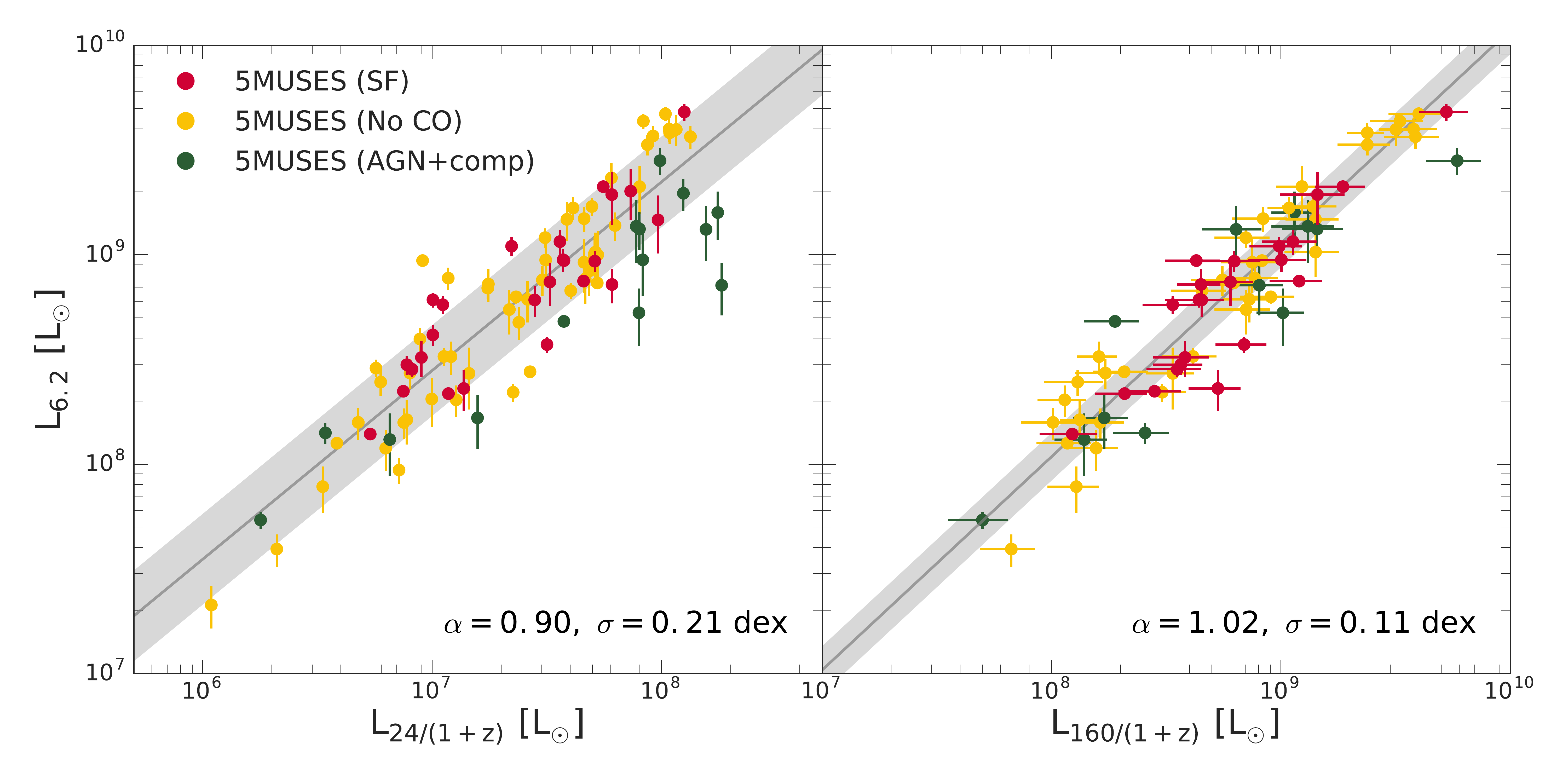}
    \caption{Correlation between \lpah\ and the 24 \si{\micro\metre} (\textit{left}) and 160 \si{\micro\metre} luminosity (\textit{right}) for the 5MUSES sample. Colour coding follow Figure \ref{fig:lir_lpah}. The dark grey line depicts the best linear regression fit to the 5MUSES sample and the $1\sigma$ dispersion of the correlation presented as the shaded region. Note the reduced intrinsic scatter of the \lpah$-L_{160}$ relation as opposed to \lpah$-L_{24}$ and the fact that AGN/composite sources are clear outliers in the \lpah$-L_{24}$ relation. The fit parameters are listed in Table \ref{tab:scaling}.}
    \label{fig:warm_combined}
\end{figure*}

\begin{figure*}
	\centering
	\includegraphics[width=\textwidth, trim={0 0 0 0}, clip]{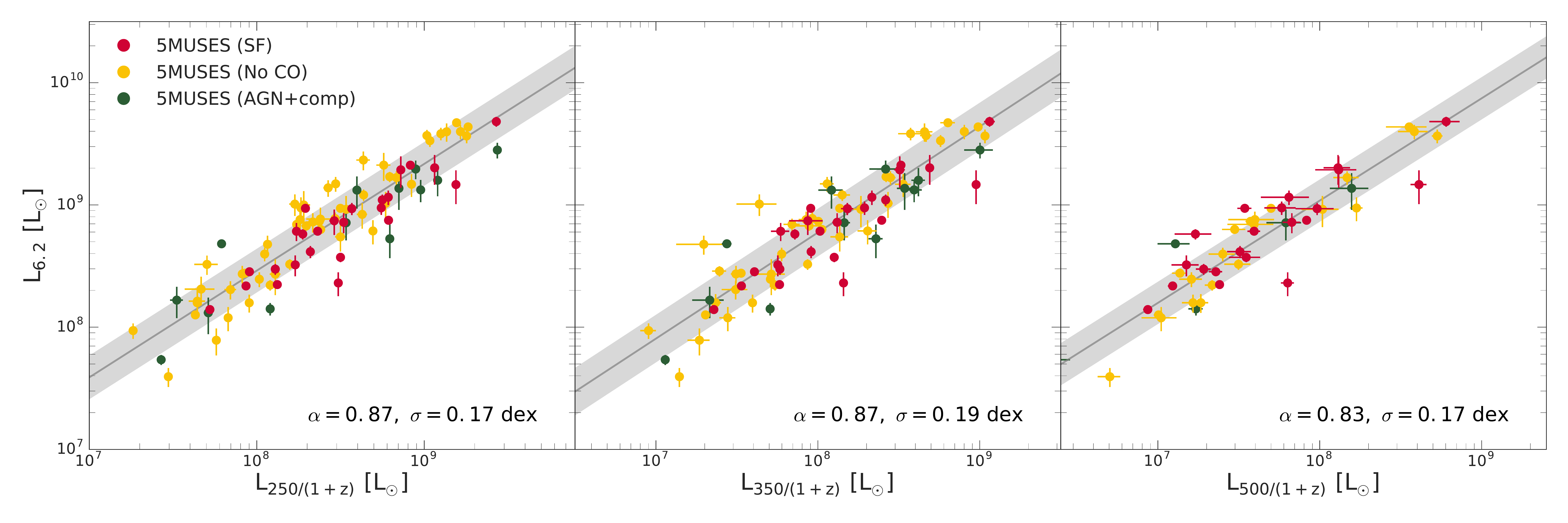}
    \caption{Correlation between \lpah\ and the 250 \si{\micro\metre} (\textit{left}), 350 \si{\micro\metre} (\textit{middle}), and 500 \si{\micro\metre} luminosity (\textit{right}) for the 5MUSES sample. The fit parameters are listed in Table \ref{tab:scaling}.}
    \label{fig:colddust_pah}
\end{figure*}

Using \s\ and \h\ photometric observations of the 5MUSES sample, we derive monochromatic luminosities at 24, 160, 250, 350 and 500 \si{\micro\metre} (see Section \ref{sec:sed}) and plot them against \lpah\ in Figure \ref{fig:warm_combined} and \ref{fig:colddust_pah}, including galaxies with secure dust luminosities ($>$3$\sigma$). To minimise the effects of K-correction, we restrict our sample to a redshift range of $0.1<z<0.3$ ($\langle z\rangle =0.12 \pm 0.05$). While \lpah\ is found to correlate with both the warm dust emission as traced by $L_{\rm 24}$ as well as with the colder dust emission (at $L_{\rm 160}$, $L_{\rm 250}$, $L_{\rm 350}$ and $L_{\rm 500}$), we obtain a lower scatter for the latter, even when AGNs are excluded from the fit.  {We note that quite naturally galaxies with the presence of an AGN appear as prominent outliers only in the \lpah$-L_{\rm 24}$ relation due to the intrinsic AGN dust emission that peaks between rest-frame 15 to 60 \si{\micro\metre} \citep{Mullaney2011}, boosting the $L_{\rm 24}$ (for fixed \lir) with respect to star-formation dominated galaxies.}

The correlation between the cold dust and PAH emission has been supported by various spatially resolved observations of local galaxies. For example, using \s\ observations of local normal galaxies, \citet{Bendo2008} find the PAH emission to be well-correlated with the 160 \si{\micro\metre} emission on spatial scales of $\sim$2 kpc, and a significant scatter in the relation between PAH and 24 \si{\micro\metre} emission, concluding that the PAHs are associated with the diffuse, cold dust. Similar results, based on SMC observations, were reached by \citet{Sandstrom2010, Sandstrom2012}, who also reported a strong correlation between the PAH and CO(1--0) emission. Furthermore, \citet{Haas2002}, find a good spatial coincidence between the 850 \si{\micro\metre} continuum emission and the strength of the PAH 7.7 \si{\micro\metre} line, suggesting again that the PAH carriers are preferentially related to the regions dominated by cold dust and molecular clouds, where they are excited mainly by the interstellar radiation field.

These findings are also in agreement with recent modeling studies of the various dust components within \hii\ regions and their surrounding envelopes \citep[Pavlyuchenkov et al. submitted; see also][]{Akimkin2015, Akimkin2017}. They argue for a lower correlation between PAHs and graphite grains, responsible for the majority of the dust emission at 24 \si{\micro\metre}, due to possible destruction of PAHs within the \hii\ regions. By modeling the intensity distributions of the different dust components, they find similar intensity distributions between PAHs and silicates that are the dominant dust component at $\sim 100-500$ \si{\micro\meter}. This could indicate that PAHs located in the molecular clouds are excited by escaping UV photons from the \hii\ regions (Pavlyuchenkov et al. submitted).

Put together, our analysis suggests that on global integrated scales PAH emission is linked to the cold dust and the CO emission in our sample, both of which are tracers of the molecular gas.\newpage


\subsection{The \texorpdfstring{\lpah--\mol}{L6.2-Mgas} relation in MS galaxies}\label{sec:mgas}
The total molecular gas mass can be estimated from the observed CO luminosity, assuming a CO--${\rm H_2}$ conversion factor, \aco: \mol [\msol]$=$\aco$\times$\lco. Corollary, the linear relation between \lpah\ and \lco\ ratio presented in the previous section, can be used to convert \lpah\ to molecular gas masses. By using the median \lpah/\lco\ ratio of $(0.37\pm0.18)$ \lsol/(\ulco) as indicated by our data, we define:
\begin{eqnarray} \label{eq:l62_mgas_lco}
M_{\rm H_{2}}\,\, [\rm M_{\sun}]= \alpha_{\rm CO} \times (2.7\pm 1.3) \times L_{\rm 6.2}
\end{eqnarray}

However, both observational and theoretical studies suggest that \aco\ varies with specific properties of the ISM, including metallicity and galaxy morphology \citep[e.g.][]{Leroy2011, Narayanan2012, Papadopoulos2012, Sandstrom2013}, ranging between  $\langle$\aco$\rangle\approx 4.5$ \uaco\ for normal MS galaxies and $\langle$\aco$\rangle \approx 0.8$ \uaco\ for local ULIRGs  and high-$z$ SBs \citep[e.g..][]{Solomon1987,Tacconi2006, Tacconi2008, Daddi2010b, Leroy2011, Magdis2011, Magdis2012, Casey2014}. 

To avoid the dependency on \aco, we derive molecular gas mass estimates using the FIR dust continuum observations. This method relies on the fact that \mgas\ can be derived from the dust mass by exploiting the well-calibrated gas-to-dust mass  ratio ($\delta_{\rm GDR}(Z)$) \citep[e.g..][]{Leroy2011, Magdis2012, Berta2016, Tacconi2018}: 
\begin{eqnarray} \label{eq:GDR_Mdust}
 \delta_{\rm GDR} \times M_{\rm dust} \equiv M_{\rm gas} = M_{\rm {H}_{\rm I}} + M_{\rm {H}_{2}}
\end{eqnarray}
\noindent where \mat\ is the atomic gas mass. Although the atomic-to-molecular gas ratio is not known at high redshift, current models suggest \mol\ dominates over \mat\ at high$-z$ and high stellar surface densities \citep[e.g..][]{BlitzRosolowsky2006, Obreschkow2009} and thus \mgas$\approx$\mol. Since the method is metallicity-dependent, and in the absence of direct metallicity estimates, we choose to restrict our sample to massive galaxies with log(\ms/\msol)$>10$ that are known to follow the mass-metallicity relation as well as the FMR relation \citep[e.g..][]{Mannucci2010} at least out to $z\sim2$. 
Moreover, whether SB systems, like local ULIRGs, follow the FMR relation or whether they are more metal-rich with respect to normal galaxies at fixed stellar mass, is still an open debate \citep[e.g..][]{Magdis2011, Magdis2012, Silverman2015, PereiraSantaella2017, Rigopoulou2018}. 
To avoid the uncertainties and systematics introduced by the metallicity of SB galaxies, we choose to omit them from our analysis in this section. Instead we focus on massive MS galaxies from the 5MUSES sample with sufficient FIR coverage (out to $\lambda_{\rm rest} >$ 250 \si{\micro\metre} to ensure robust \md\ estimates.

\begin{figure}
	\centering
	\includegraphics[width=.47\textwidth, trim={0 0 0 0}, clip]{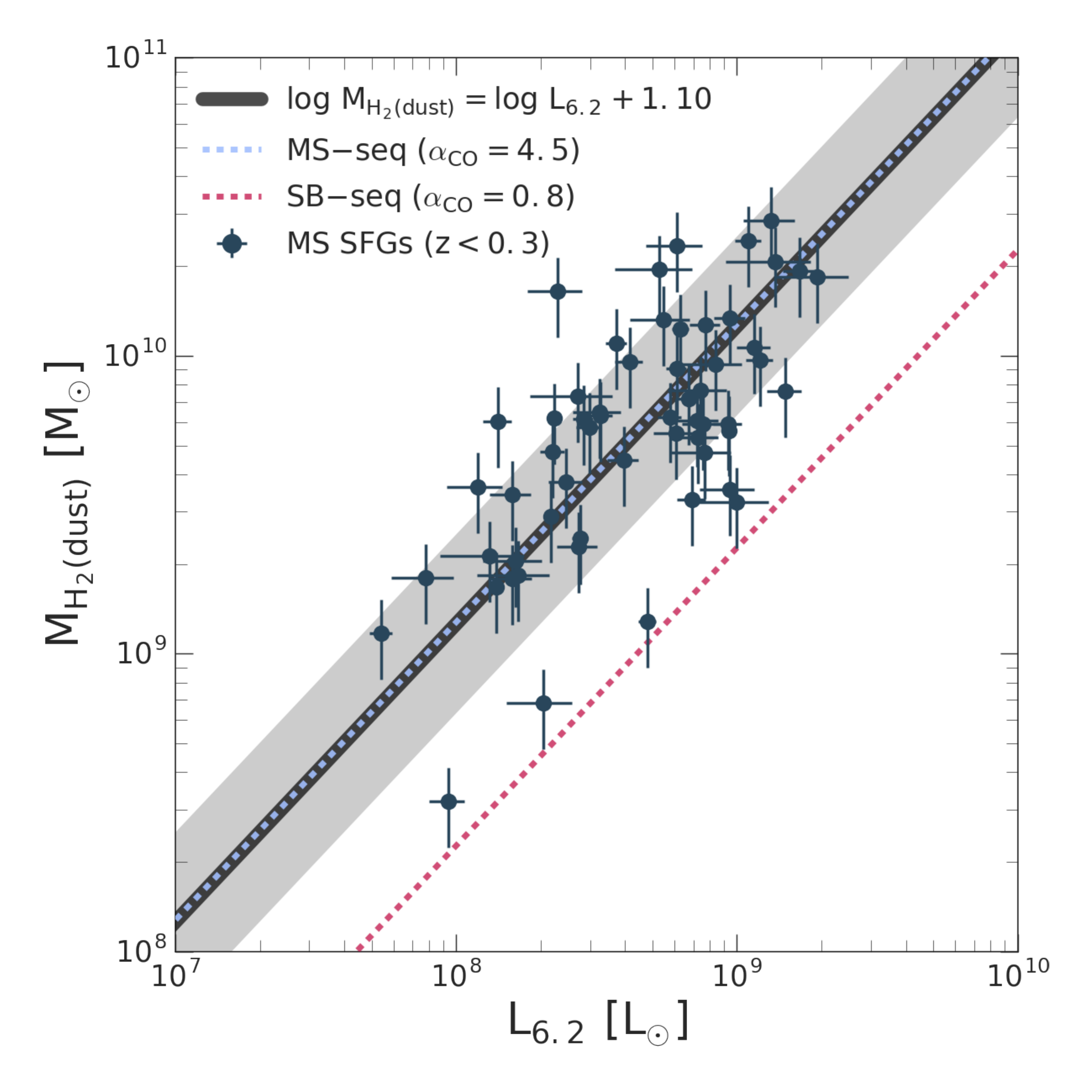}
    \caption{Correlation between \lpah\ and dust-derived molecular gas mass (\mol) for normal galaxies assuming solar metallicity. The sample includes 5MUSES MS galaxies at $z < 0.3$ with log(\ms/\msol)$>10$. The black line corresponds to the best fit with a fixed slope of 1 with an intrinsic scatter of 0.30 dex (grey shaded region). Light blue and red dashed lines are the \mol--\lco\ relation for MS and SB galaxies adopting \aco$=4.5$ \uaco\ and \aco$=0.8$ \uaco, respectively.}
    \label{fig:l_62_mgas}
\end{figure}

Molecular gas masses are then inferred using the $\delta_{\rm GDR}-Z$ metallicity relation: ${\rm log}~ \delta_{\rm GDR}= (10.54 \pm 1.0) - (0.99 \pm 0.12) \times (12 + {\rm log}({\rm O}/{\rm H}))$ from \citet{Magdis2012}. The resulting \mol\ estimates versus \lpah\ are presented in \ref{fig:l_62_mgas}, yielding: 

\begin{eqnarray} \label{eq:l62_mgas}
\mathrm{log} \left( \frac{M_{\rm H_{2}}}{M_{\odot}} \right) = {\rm log}  \left( \frac{L_{6.2}} {L_{\sun}} \right) + (1.10 \pm 0.02) 
\end{eqnarray}

\noindent with an intrinsic scatter of $\sigma$ = 0.28\,dex. As a sanity check, we also overplot the \mol--\lpah\ relation using equation 3, adopting \aco$=4.5$ \uaco, a typical value for normal SFGs. Indeed, we see that this relation is in excellent agreement with our data, reassuring that both the dust-based \mol\ estimates and the \lco-based \mol\ estimates with \aco$=4.5$ \uaco\ are consistent. Finally, we overplot equation 3 assuming \aco$=0.8$ \uaco\ to indicate the expected location of SBs in Figure \ref{fig:l_62_mgas}. An attempt to explain the physical origin of the derived \lpah--\mol\ relation and a discussion of its importance and limitations are presented in the following Section.

\section{Discussion}
In the previous sections, we have explored the global scaling laws between \lpah, \lco, and \lir\ for a large  compilation of different galaxy populations at various redshifts. We have seen that variations in the \lpah/\lir\ and \lco/\lir\ ratios among different galaxy populations are not permeated in their \lpah/\lco\ ratios that instead appear to be rather constant through a universal \lpah--\lco\ relation with a slope of unity. 

\begin{figure}
	\centering
	\includegraphics[width=.47\textwidth, trim={0 0 0 0}, clip]{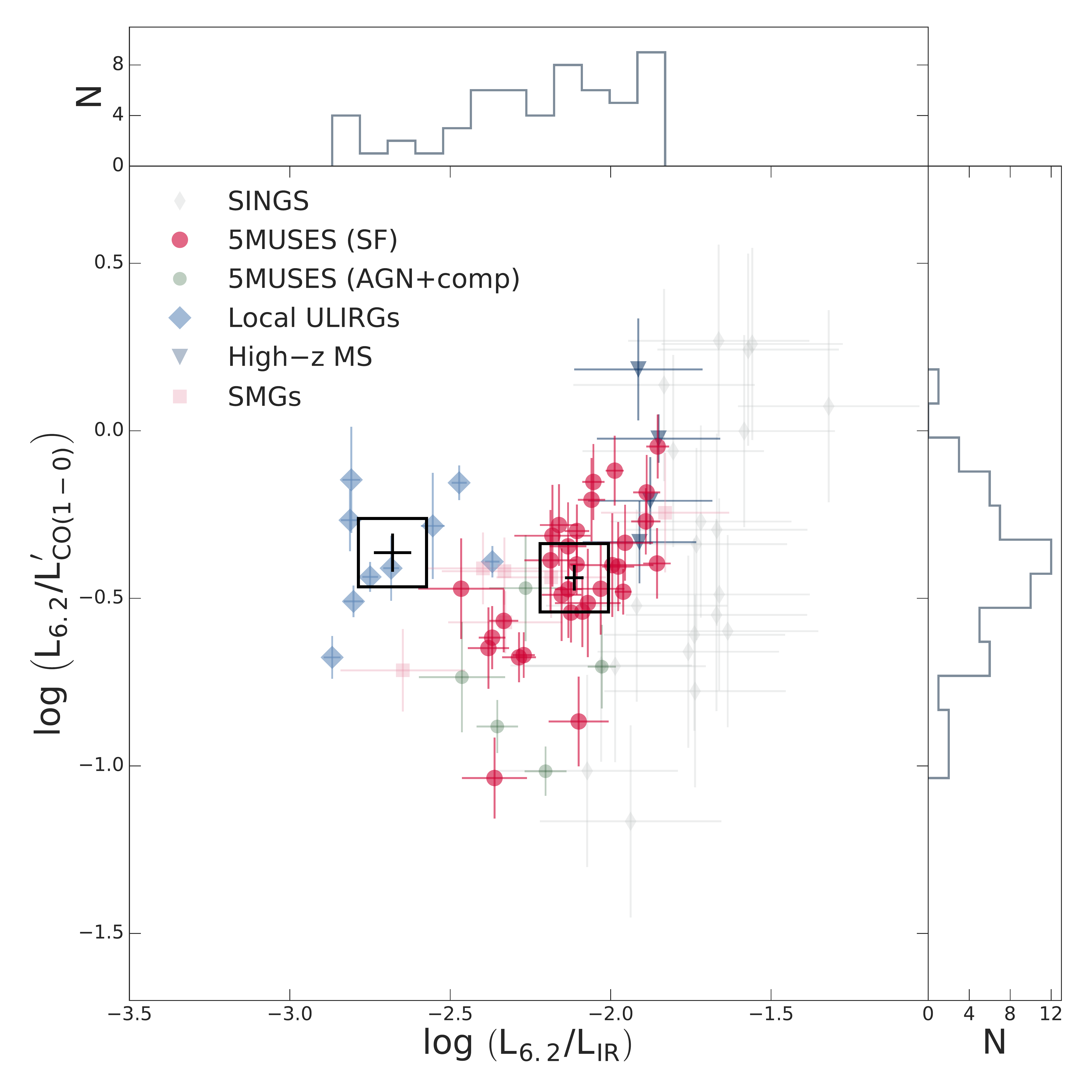}
    \caption{log(\lpah/\lco) vs. log(\lpah/\lir) for the full sample with both PAH and CO emission. Colour coding and symbols follow Figure \ref{fig:lir_lpah}. The black open squares correspond to the median values for the 5MUSES SFGs and the local ULIRGs. The top and right panels show the distribution of \lpah/\lir\ and \lpah/\lco, respectively.}
    \label{fig:comparison}
\end{figure}

This is further demonstrated in Figure \ref{fig:comparison}. While the distribution of the log(\lpah/\lco) ratios is symmetric around the central value, we find a negatively skewed distribution of the log(\lpah/\lir) ratios. This skewness is driven by systematically lower log(\lpah/\lir) ratios of local ULIRGs. In fact, we find a $8.70\sigma$ significant difference between the mean values of log(\lpah/\lir) for the local ULIRGs ($-2.68\pm 0.06$) and the 5MUSES SFGs ($-2.11\pm 0.03$). On the other hand, we do not find any significant systematic variations in the log(\lpah/\lco) ratios (ULIRGs: $0.36\pm0.06$, 5MUSES SFGs: $0.44\pm0.04$). This suggests that \lpah\ is, in fact, a better tracer of CO (and thus of \mol) than the total IR emission.

While the emerging PAH--CO luminosity relation has also been reported in spatially resolved observation of local galaxies, its physical origin (if any) is still debated. \citet{Bendo2010} suggest that the correlation between PAH and CO emission observed in NGC 2403 on large scales can be explained if they share similar excitation mechanisms, or if the molecular cloud formation is triggered in regions with stellar potential wells as described in \citet{Leroy2008}. For the latter, CO emission could then arise from newly formed molecular clouds, while starlight in the potential wells would heat surrounding regions enhancing the PAH emission. This scenario would also explain the variations between PAH and CO emission on sub-kpc scales and the similar radial profiles on larger spatial scales \citep{Regan2006}. Finally, recent studies have suggested that in dense PDRs, PAHs may be responsible for a significant fraction of the $\rm{H}_{2}$ formation at a rate comparable to that of $\rm{H}_{2}$ formation on dust grains \citep{Castellanos2018a,Castellanos2018b}. If so, a correlation between PAHs and \mol\ could thus be expected.

\begin{figure}
	\centering
	\includegraphics[width=.5\textwidth, trim={100 0 80 50}, clip]{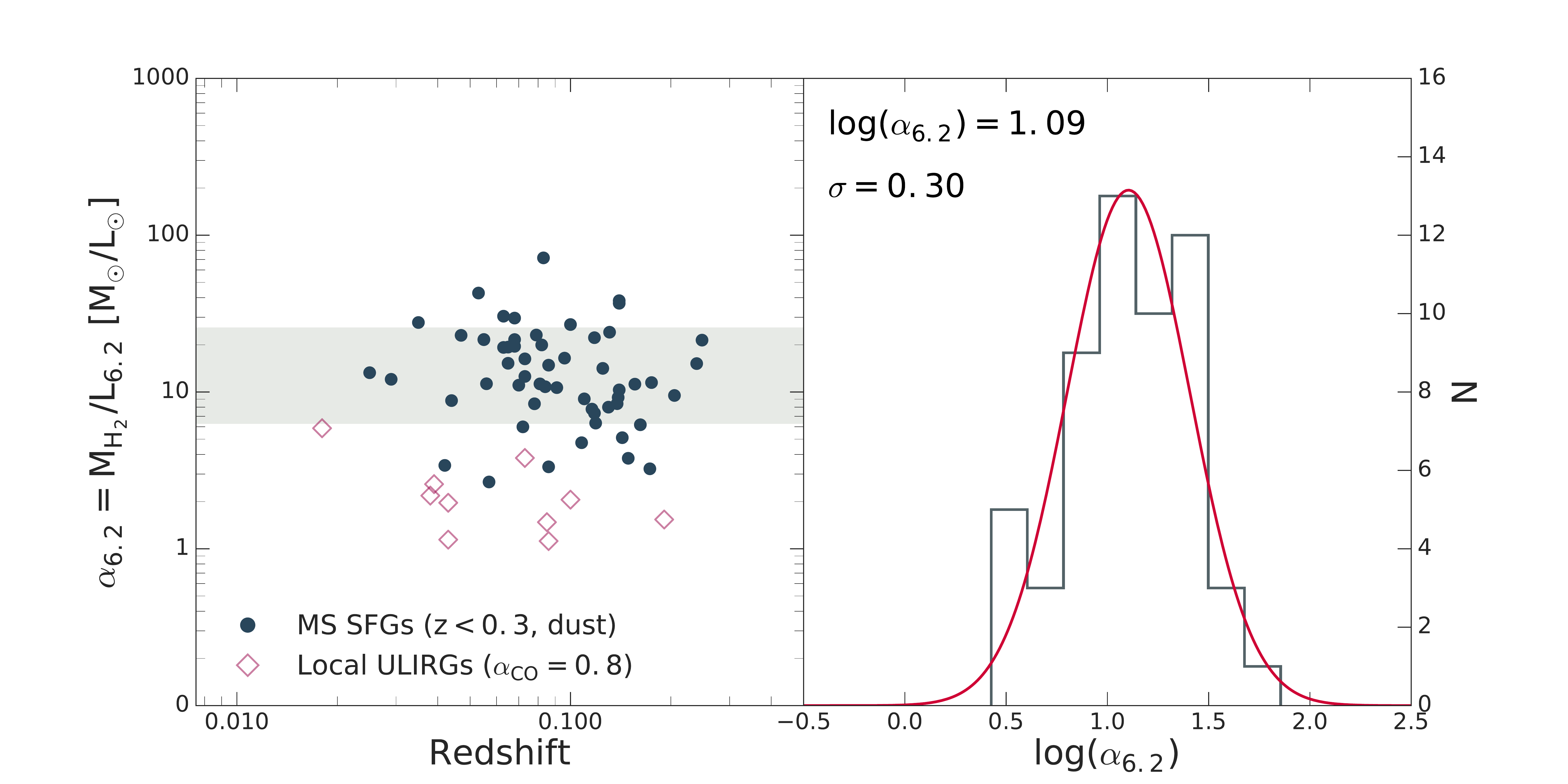}
    \caption{\textit{Left}: Correlation between the \apah\ conversion factor vs. redshift. The sample includes dust-derived \mgas\ estimates of 5MUSES MS galaxies at $z<0.3$ (dark blue). The grey region depicts the average \apah\ value of the sample and the $1\sigma$ dispersion. We overplot local ULIRGs (red) assuming \aco$=0.8$ \uaco. \textit{Right}: Distribution of log(\apah) for the 5MUSES MS galaxies. The red line corresponds to the 
    best-fit Gaussian profile to the data with a mean value of 1.09 and a standard deviation $\sigma =$ 0.30.}
    \label{fig:alpha_62}
\end{figure}

The CO--PAH luminosity relation is not the only piece of evidence for a potential link between \lpah\ and \mol. In Section \ref{sec:mgas}, we used CO-independent \mol\ estimates through the $\delta_{\rm GDR}-Z$ technique and found a linear \lpah--\mol\ relation for normal SFGs. The relation in eq. \ref{eq:l62_mgas} can then be used to define: \mol$=$\apah $\times$\lpah\, where \apah\ is the parameter converting \lpah\ to \mol. The inferred \apah\ conversion factor of our sample of MS galaxies are shown in Figure \ref{fig:alpha_62} (\textit{left}). It appears that \apah\ is independent of redshift for MS galaxies with an average value of $\langle {\rm log}(\alpha_{\rm 6.2}) \rangle=1.09$ and a standard deviation of 0.30 (Figure \ref{fig:alpha_62} \textit{right}). An obvious caveat for the derivation of \mol\ from \lpah\, similar to the $\delta_{\rm GDR}$ and CO technique, is the dependence on metallicity which becomes challenging especially for SB systems and low-mass galaxies. For speculation, we overplot local ULIRGs in Figure \ref{fig:alpha_62} (\textit{left}) inferring their \mol\ estimates from \lco\ and assuming the commonly adopted \aco $=0.8$ \uaco. Naturally, lower \aco\ values lead to lower \apah\ values for SB systems. 

With the upcoming launch of the {\it James Webb Space Telescope} ({\it JWST}), PAH features will be detected and spatially resolved with MIRI out to $z \sim$3.5. Our analysis suggests that PAHs can be used as a tool to infer the total amount and the spatial distribution of the molecular gas in systems that will probably be too faint to be detected by \h\ and thus lack any FIR photometric coverage. Since the most prominent and bright PAH feature is the one at 7.7 \si{\micro\metre}, it is worth to present its scaling relation with \mol\ as well. Repeating our analysis using \lpahsev, we then find:  

\begin{eqnarray} \label{eq:l76_Mgas}
{\rm log} \left( \frac{M_{\rm H_{2}}}{M_{\sun}} \right) =  \mathrm{log}  \left( \frac{L_{\rm 7.7}} {L_{\sun}} \right) + (0.55 \pm 0.02)
\end{eqnarray}
with an intrinsic dispersion of $0.28$ dex.

\section{Conclusions}
We have presented 24 new CO(1--0) observations of intermediate redshift ($0.03<z<0.30$) SFGs drawn from the 5MUSES sample that also benefits from existing \s\ IRS spectroscopy and FIR photometry observations from \s\ and \h. Complementing our study with literature CO, PAH, and dust observations, we investigate scaling relations between the various components of the ISM of galaxies covering a wide range of redshifts and physical conditions. We summarise our conclusions as follows:

\begin{itemize}[leftmargin=.5cm]
\item We confirm the existence of a correlation between the PAH and CO emission on global integrated scales and for the first time determine its slope and scatter in a robust statistical way. The linear and tight \lpah--\lco\ correlation (slope of $\sim 1.0$ and intrinsic scatter $\sigma=0.26$ dex) is followed by the majority of galaxies at all redshifts, independent of the galaxy type.

\item We find evidence that, on galaxy integrated scales, \lpah\ traces better the cold dust ($\lambda > 100$ \si{\micro\metre}) rather than the warm dust emission ($\lambda = 24$ \si{\micro\metre}). This is in agreement with spatially resolved observations of local galaxies. The fact that both CO and cold dust emission are tracing molecular gas, motivates us to propose that PAHs may serve as a gas tracer in SFGs. 

\item We define a \apah\ = \mol/\lpah\ conversion factor of 2.7 $\times$ \aco, where \aco\ is the \lco\ to \mol\ conversion factor. For normal SFGs we find \apah\ = 12.30 \msol/\lsol\ ($\sigma =$ 0.30\,dex), which is consistent with \aco\ $\approx$ 4.5 \uaco, typical of normal SFGs.
\end{itemize}

We conclude that \lpah\ can effectively probe the molecular gas mass in galaxies within a factor of $\sim2$ and propose that with the launch of JWST, PAHs could possibly serve as a useful tool to trace the ISM properties in SFGs up to $z\sim3.5$.
 
\section{Acknowledgements}
IC acknowledges support from Villum Fonden research grant (13160). GEM acknowledges support from the Carlsberg Foundation and a research grant (13160) from Villum Fonden. GEM, ST, CGC, and MS acknowledge support from the ERC Consolidator Grant funding scheme (project ConTExt, grant number No. 648179. This work includes observations carried with the IRAM 30m telescope, which is supported by INSU/CNRS (France), MPG (Germany) and IGN (Spain). The Cosmic Dawn Center is funded by the Danish National Research Foundation.

\section{Supporting information}
Supplementary data are available at MNRAS online.

\onecolumn
\tiny
\renewcommand*{\arraystretch}{1.4}
\begin{longtable}{ccccccccccc}
\caption{General properties of the 5MUSES sample.}\\
\hline
\hline
$\mathrm{ID}^{\mathrm{a}}$ & R.A. & Decl. & $z$ & log(\lpah/\lsol)$^{\rm b}$ & ${EW}_{\rm 6.2}$ & log(\lpahsev/\lsol)$^{\rm b}$ & log(\Mstar/\msol)$^{\mathrm{c}}$ & log(\lir/\lsol) & log(\md/\msol) & log(\mol/\msol)  \\
& [hh:mm:ss] & [dd:mm:ss] &  &  & [\si{\micro\m}] & & & & & \\
\hline
\label{tab:muses_all}
2   & 02:15:03.5 & --04:24:21.7   & 0.137 & $8.58\pm0.19$  & $0.78\pm0.01$  & $9.19\pm0.14$  & 10.51 & $10.86\pm0.04$ & $8.01\pm0.28$ & $10.01$ \\
4   & 02:15:57.1 & --03:37:29.1   & 0.032 & $7.59\pm0.08$  & $0.50\pm0.05$   & $8.09\pm0.09$  & 9.97  & $9.83\pm0.02$  & $7.18\pm0.29$ & $9.18$  \\
5   & 02:16:38.2 & --04:22:50.9   & 0.304 & $9.06\pm0.24$  & \textless0.09 & $9.48\pm0.11$  & 10.80  & $11.54\pm0.02^{\rm d}$ & --            & --             \\
6   & 02:16:40.7 & --04:44:05.1   & 0.870  & $9.13\pm1.62$  & \textless0.05 & --             & 11.82 & $12.70\pm0.01^{\rm d}$  & --            & --             \\
8   & 02:16:49.7 & --04:25:54.8   & 0.143 & $9.01\pm0.09$  & $1.11\pm0.06$  & $9.48\pm0.08$  & 10.06 & $10.99\pm0.01$ & $7.29\pm0.42$ & $9.29$  \\
... & ... & ... & ... & ... & ... & ... & ... & ... & ... & ...
 \\ 
\hline
\label{tab:muses_full}
\end{longtable}
\begin{flushleft}
NOTE -- Col. (1): Source name; col. (2): Right ascension in units of hours, minutes, and seconds; col. (3): Declination in units of degrees, arcminutes, and arcseconds; col. (4): Redshift; col. (5): PAH 6.2 \si{\micro\metre} luminosity; col. (6): Equivalent width of the PAH 6.2 \si{\micro\metre} feature; col. (7): PAH 7.7 \si{\micro\metre} luminosity; col. (8): Stellar mass; col. (9): Infrared luminosity (integrated from $8-1000$ \si{\micro\metre}); col. (10): Dust mass; col. (13): Molecular gas mass derived using the \md--\gdr\ method.\\
$^{\rm a}$5MUSES ID name.\\
$^{\rm b}$PAH luminosity from \citet{Magdis2013}.\\
$^{\rm c}$Stellar masses from \citet{Shi2011}.\\
$^{\rm d}$ \lir\ from \citet{Shi2011}.\\
\end{flushleft}

\begin{longtable}{lllllllllllll} 
\caption{Galaxies from the literature with PAH, IR and/or CO.}\\
\hline
\hline
Name            & R.A.         & Decl.        & Sample     & $z$     & log(\lpah/\lsol) & \ew\ & log(\lpahsev/\lsol) & log(\lir/\lsol) & Line    & log(\lco/\lsol) \\
 & [hh:mm:ss] & [dd:mm:ss] & &  &  & [\si{\micro\meter}] & & &  & & \\
\hline
NGC 3049                & 09:54:49.59   & +09:16:18.1   & SINGS           & 0.006 & $7.81\pm0.02^{\rm a}$    & --   &  --       & $9.65\pm0.02$  & CO(3--2)      & $7.68\pm 0.21$              \\
IRAS 10565+2448         & 10:59:18.1    & +24:32:34      & local ULIRGs  & 0.043 & $9.57\pm0.01$    & --  &  --        & $12.04\pm0.02^{\rm c}$  & CO(1--0) & $9.72\pm0.03^{\rm c}$   \\
GN26   & 12:36:34.51   & +62:12:40.9   & SMGs             & 1.223 & $10.48\pm0.02$   & $0.38\pm0.04$   & $11.01\pm0.10$      & $12.66\pm0.17^{\rm c}$  & CO(2--1) & $10.92\pm0.12^{\rm c}$    \\
GN70.211  & 12:37:10.60   & +62:22:34.5   & High-$z$ SFGs             & 1.523 & $10.39\pm0.10$   & --   & $10.84\pm0.12$   & $11.94\pm0.17^{\rm c}$  & CO(1--0) & $10.20\pm0.12^{\rm c}$   \\
MIPS506   & 17:11:38.59  & +58:38:38.58  & High-$z$ SBs       & 2.470  & --  & $0.30\pm0.14^{\rm d}$ & $11.11\pm0.13$ &  $12.93\pm0.09^{\rm e}$  & CO(3--2) & $10.52\pm0.09^{\rm e}$    \\
... & ... & ... & ... & ... & ... & ... & ... & ... & ... \\
\hline
\label{tab:literature}
\end{longtable}
\begin{flushleft}
NOTE -- Col. (1): Source name; col. (2): Right ascension in units of hours, minutes, and seconds; col. (3): Declination in units of degrees, arcminutes, and arcseconds; col. (4): Galaxy sample or selection; col. (5): Redshift; col. (6): PAH 6.2 \si{\micro\metre} luminosity; col. (7): Equivalent width of the PAH 6.2 \si{\micro\metre} feature; col. (8): Infrared luminosity (integrated from $8-1000$ \si{\micro\metre}); col. (9): Observed CO line; col. (10): CO(1--0) luminosity.\\
$^{\rm a}$ From \citet{Smith2007}.\\
$^{\rm b}$ From \citet{Wilson2012}. Converted to CO(1--0) using $r_{32/10}=0.18\pm0.02$.\\
$^{\rm c}$ From \citet{Pope2013}. For the CO luminosities, we corrected higher-$J$ transitions using the conversion scheme listed in \citet{Bothwell2013}: $r_{21/10}=0.84\pm0.13$, $r_{32/10}=0.52\pm0.09$, $r_{43/10}=0.41\pm0.07$.\\
$^{\rm d}$ From \citet{Sajina2007}.\\
$^{\rm e}$ From \citet{Yan2010}.\\
$^{\rm f}$ From \citet{Pope2008b}.\\
$^{\rm g}$ From \citet{Kirkpatrick2014}.\\
\end{flushleft}

\begin{table*}
\centering
\caption{Linear scaling relations between the emission from PAHs, IR, and CO, and various galaxy properties.}
\label{tab:scaling}
\begin{tabular}{llcccc} 
\hline
\hline
log $x$ & log $y$ & $\alpha$ & $\beta$ & $\sigma$ & Sample \\
\hline
\lir\ [\lsol] & \lpah\ [\lsol] & $0.98 \pm 0.03$ & $-1.89 \pm 0.30$ & $0.13$ & SFGs \\ 
\lir\ [\lsol] & \lpah\ [\lsol] & $0.81 \pm 0.03$ & $-0.04 \pm 0.29$ & $0.21$ & All \\ 

\lir\ [\lsol] & \lpahsev\ [\lsol] & $1.00 \pm 0.03$ & $-1.53 \pm 0.28$ & $0.13$ & SFGs \\ 
\lir\ [\lsol] & \lpahsev\ [\lsol] & $0.84 \pm 0.02$ & $-0.15 \pm 0.26$ & $0.20$ & All \\ 

\lco\ [\ulco] & \lpah\ [\lsol] & $1.02 \pm 0.06$ & $-0.65\pm0.31$ & $0.24$ & SFGs \\ 
\lco\ [\ulco] & \lpah\ [\lsol] & $1.03 \pm 0.06$ & $-0.73\pm0.38$ & $0.26$ & All \\ 
\lco\ [\ulco] & \lpah\ [\lsol] & $0.99 \pm 0.07$ & $-0.37\pm0.70$ & $0.24$ & With CO(1--0) \\  

\lco\ [\ulco] & \lpahsev\ [\lsol] & $1.03 \pm 0.08$ & $-0.13\pm 0.19$ & $0.21$ & SFGs \\ 
\lco\ [\ulco] & \lpahsev\ [\lsol] & $1.04 \pm 0.08$ & $-0.21\pm 0.20$ & $0.23$ & All \\ \hline

$L_{24}$ [\lsol] & \lpah\ [\lsol] & $0.99 \pm 0.05$ & $1.51 \pm 0.37$ & $0.20$ & 5MUSES SFGs \\ 
$L_{24}$ [\lsol] & \lpah\ [\lsol] & $0.90 \pm 0.05$ & $2.14 \pm 0.35$ & $0.21$ & 5MUSES \\ 

$L_{160}$ [\lsol] & \lpah\ [\lsol]& $1.02 \pm 0.04$ & $-0.11 \pm 0.37$ & $0.11$ & 5MUSES \\ 
$L_{250}$ [\lsol] & \lpah\ [\lsol] & $0.87 \pm 0.04$ & $1.48 \pm 0.31$ & $0.17$ & 5MUSES \\ 
$L_{350}$ [\lsol] & \lpah\ [\lsol]& $0.87 \pm 0.05$ & $1.83 \pm 0.38$ & $0.19$ & 5MUSES \\ 
$L_{500}$ [\lsol] & \lpah\ [\lsol] & $0.83 \pm 0.06$ & $2.34 \pm 0.42$ & $0.17$ & 5MUSES  \\ 
\lpah\ [\lsol] & \lpahsev\ [\lsol] & $1.00 \pm 0.02$ & $0.62 \pm 0.13$ & $0.01$ & 5MUSES  \\ \hline
$L_{24}$ [\lsol] & \lpahsev\ [\lsol] & $0.97 \pm 0.05$ & $2.15 \pm 0.37$ & $0.21$ & 5MUSES SFGs \\ 
$L_{24}$ [\lsol] & \lpahsev\ [\lsol] & $0.79 \pm 0.05$ & $3.46 \pm 0.40$ & $0.28$ & 5MUSES \\ 
$L_{160}$ [\lsol] & \lpahsev\ [\lsol]& $1.04 \pm 0.04$ & $0.23 \pm 0.37$ & $0.12$ & 5MUSES \\ 
$L_{250}$ [\lsol] & \lpahsev\ [\lsol] & $0.87 \pm 0.04$ & $2.03 \pm 0.3$ & $0.19$ & 5MUSES \\ 
$L_{350}$ [\lsol] & \lpahsev\ [\lsol]& $0.85 \pm 0.05$ & $2.45 \pm 0.40$ & $0.22$ & 5MUSES \\ 
$L_{500}$ [\lsol] & \lpahsev\ [\lsol] & $0.82 \pm 0.06$ & $3.00 \pm 0.48$ & $0.22$ & 5MUSES \\ \hline
\lpah\ [\lsol] & \lpahsev\ [\lsol] & $1.00 \pm 0.02$ & $0.62 \pm 0.13$ & $0.01$ & 5MUSES \\
\lpah\ [\lsol] & \lpahsev\ [\lsol] & $1.02 \pm 0.02$ & $0.36 \pm 0.19$ & $0.06$ & All \\ \hline
\lpah\ [\lsol] & \mol$^{\rm a}$ [\msol] & 1.00 (fixed) & $1.10 \pm 0.01$ & $0.28$ & 5MUSES MS SFGs$^{\rm b}$ \\ 
\lpahsev\ [\lsol] & \mol$^{\rm a}$ [\msol] & $1.00$ (fixed) & $0.55 \pm 0.02$ & $0.28$ & 5MUSES MS SFGs$^{\rm b}$ \\  
\hline \hline
\end{tabular}
\begin{flushleft}
  \textbf{Notes} -- The linear fits are obtained in the logarithmic space : ${\rm log}~y = 
  \alpha \times {\rm log}~x + \beta$. The best-fit parameters and the intrinsic scatter are estimated 
  from the Bayesian linear regression method described in \citet{Kelly2007}. \\
  $^{\rm a}$Dust-derived molecular gas masses assuming solar metallicity.\\
  $^{\rm b}$With log(\ms/\msol)$>10$.
\end{flushleft}
\end{table*}



\bibliographystyle{mnras}
\bibliography{PAHs}




\bsp	
\label{lastpage}
\end{document}